\documentclass[pra,twocolumn,showpacs]{revtex4}

\usepackage{graphicx}
\usepackage{amsfonts}
\usepackage{amsmath}
\usepackage{amssymb}

\usepackage{amstext} 
\usepackage{slashed}
\usepackage{dsfont}

\begin{document}

\title{Bound-free $\mathbf{e}^+ \mathbf{e}^-$ pair creation with a linearly polarized laser field and a nuclear field}

\author{C.\ Deneke}
\author{C.\ M\"{u}ller}

\affiliation{Max-Planck-Institut f\"{u}r Kernphysik, Saupfercheckweg 1, 69117 Heidelberg, Germany}

\begin{abstract}

The process of bound-free pair production of electrons and positrons in combined laser and Coulomb fields is investigated. It is assumed that an ion collides at relativistic speed with an intense x-ray laser beam of linear polarization. The process proceeds nonlinearly due to simultaneous absorption of a few laser photons. The capture of the electron into the ground state and the $L$-shell is considered.
The scaling of the total rate, the angular distributions of the emitted positrons and a comparison to the competing free-free channel are surveyed. Numerical results of pair production rates for parameters  for the planned x-ray free electron lasers at DESY and SLAC are presented. We find that pair production with these laser facilities can become observable in the near future.

\end{abstract}

\maketitle

%
\section{Introduction}
Theoretical and experimental study of electron-positron pair creation in strong external fields is important for understanding the structure of the QED vacuum. In the past, pair production in relativistic heavy ion collisions was studied in detail \cite{Bertulani05,Baur07,Eichler95}.
With the progress of laser technology, pair production via absorption of several real photons becomes increasingly interesting. There are various schemes of pair production in laser fields, for example the production with counter-propagating laser fields (e.g. \cite{Brezin70,Salamin06,Maquet02,Mourou06,Marklund06} and references therein). However, for the observation it is necessary to come close to the critical electrical field strength $\mathcal E_c = 1.3 \times 10^{16} \mathrm{V/cm}$ \cite{Schwinger51} which is still far away from experimental realisation.

Another way is to consider pair production in combined laser and Coulomb fields. The Coulomb field can originate from a moving ion. If the ion counterpropagates the laser beam at high speed, the laser's electric field strength and photon frequency are Doppler shifted by the factor $\approx 2 \gamma$, where $\gamma$ is the relativistic Lorentz factor of the ion. 
One usually distinguishes between the multiphoton and the tunneling regime of pair production. 
In the latter, the laser field strength is close to the critical one and pair creation proceeds via the simultaneous absorption of a very large number of low frequency photons ($\sim 10^6$). 
An important quantity which arises in strong-field physics is the intensity parameter, $\xi = \frac{e \mathcal E_0}{m c \omega}$, with the laser frequency $\omega$ and peak field strength $\mathcal E_0$. Moreover, $e$, $m$ and $c$ denote the elementary charge, electron mass and speed of light, respectively. The tunneling regime corresponds to $\xi \gg 1$, thus to a high intensity and low frequency laser field.
Contrary, in the multiphoton regime the field strength is smaller, but the photon energies are sufficiently high to allow pair creation via the absorption of only a few photons. Therefore, for the multiphoton regime $\xi \ll 1$.
When electron-positron pairs are created in the vicinity of a nucleus, the electrons can be created either free or in a bound state. One refers usually to \textit{free-free} and \textit{bound-free} pair production, respectively. 
In the present work we consider bound-free pair creation in the multiphoton regime due to the head-on collision of a very fast bare ion or proton with a very intense X-ray laser field of linear polarization.

X-ray free electron lasers (XFEL) currently under construction at DESY (Hamburg, Germany) and SLAC (Standford, California) are predicted to obtain photon energies of $\hbar \omega \sim 4- 12$ keV and intensities of $I \sim 10^{18} \mathrm {W/cm^2}$ \cite{Desy02}. 
Energy conservation demands $n \hbar \omega > 2 m c^2$, the energy of $n$ absorbed laser photons has to be greater than the rest masses of the produced particles. 
Hence, together with acceleration of ions to a relativistic $\gamma$-factor of around $50$, pair creation with two or three photons from an XFEL is possible as a nonlinear process.

In 1997, the process of laser-induced pair creation has been observed at SLAC \cite{Burke97,Reiss62,Ritus72}. A highly relativistic electron beam collided with a highly intense laser beam. Energy conservation allowed the production of electron-positron pairs by absorption of five photons. Two different mechanisms were found to be responsible for the process. On the one hand the Breit-Wheeler process where in a first step one or two photons are absorbed to create a highly energetic photon via Compton scattering. In a second step this $\gamma$ photon interacts with additional laser photons to create pairs. Alternatively, pair production happens via the nonlinear Bethe-Heitler process in a single step by simultaneous absorption of several real photons in the Coulomb field of the electrons to create pairs. In the analysis of the experiment it was found that the first process dominated over the second. 
In contrast, when the laser beam collides with heavy ions instead of electrons, the high masses of the ions suppress the Breit-Wheeler process substantially and the (nonlinear) Bethe-Heitler process plays the dominant role.
Bound-free pair creation has not yet been observed in laser-ion collisions, but in a similar way bound electrons have been detected in an experiment with relativistic heavy-ion collisions \cite{Belkacem93,Belkacem94,Momberger87}.  For small nuclear charge numbers $Z$ bound-free pair creation is suppressed, but it is competitive with the free-free channel at intermediate and high $Z$ \cite{Eichler95}. The two competing channels of free-free and bound-free pair creation could be distinguished experimentally \cite{Belkacem93,Belkacem94}.


Theoretically the process of free pair creation by combined Coulomb and laser fields has been investigated by several authors \cite{Yakovlev66,Mittleman87,Roshchupkin96,Dietz98,Roshchupkin01,Avetissian03,Muller03a,Muller03c,Muller04,Kaminski06,Krajewska06,Sieczka06,Milstein06,Kuchiev07,Krajewska08} with a focus on the tunneling regime. 
Also muon pair creation in laser-ion collisions has been calculated recently \cite{Kuchiev07b,Muller08}.
Bound-free electron-positron pair creation with single high-energy photons has been surveyed before (see, e.g., \cite{Sauter31,Agger97}), but for the nonlinear case, i.e. the absorption of more than one photon, only a few calculations exist \cite{Muller03b,Matveev05}. Both consider K-shell capture of the electron and a circularly polarized laser beam \cite{Muller03b} or an ultrashort, single-cycle electromagnetic pulse \cite{Matveev05}, respectively.

It is worth mentioning that in relativistic laser-ion or laser-electron collisions also other QED processes can take place which arise from the nonlinear response of the vacuum due to virtual electron-positron pairs.
This can lead, for example, to laser photon merging \cite{Piazza08} or the emission of Unruh radiation \cite{Schutzhold06,Schutzhold08}. Further nonlinear vacuum effects in strong laser fields comprise photon-photon scattering \cite{Piazza05,Lundstrom06,Fedotov07}, photon splitting \cite{Piazza07,Brodin07} and changes in the refractive index \cite{Piazza06,Heinzl06}. The creation of bound states in a supercritical ionic field was treated in \cite{Krekora05}.

In the present paper, we consider the bound-free channel of electron-positron pair production in XFEL-nucleus collisions. In contrast to the earlier treatments of this problem \cite{Muller03b,Matveev05}, a laser beam of linear polarization is chosen.
On the one hand this choice represents the more easily achievable case for experimental realization \cite{Desy02} but leads on the other hand to more difficult calculations due to the appearance of the generalized Bessel functions (see Eq. (\ref{genbessel})). Additionally, we consider not only the capture of the electron into the ground state, but also into excited states in the $L$-shell. This gives rise to corrections to the total bound-free production channel of $\sim 15 - 20\%$. 

The paper is organized as follows: Sec. \ref{Theoretical framework} describes the theoretical framework and the analytic calculations performed to obtain the result for the differential production rates. In Sec. \ref{Results} numerical results for total and differential pair production rates are shown and discussed.
In Sec. \ref{Conclusion} a summary and conclusion are given. 

Relativistic units are used such that $c=\hbar=1$ and the elementary charge $e=\sqrt{\alpha}$ with the fine-structure constant $\alpha$.
The (scalar) four product is denoted, e.g., with $\left( p x \right)$ and $\slashed p$ means $p^{\mu} \gamma_{\mu}$. The standard metric $\left(+---\right)$ and Dirac matrices $\gamma_{\mu}$ are employed \cite{Bjorken64}.
%
\section{Theoretical framework} \label{Theoretical framework}

The process of bound-free pair production in combined laser and Coulomb fields can be described within an S-matrix formalism. The amplitude for the transition from the negative energy continuum to a bound state reads \cite{Muller03b}

\begin{equation} \label{s-matrix}
S_{\mathrm{post}} = - i e \int d^4x \overline{\phi} \slashed{A}_L \Psi^{\left( +\right) } \,\,.
\end{equation} 

Eq. (\ref{s-matrix}) is the so called \textit{post} form (in contrast to the \textit{prior} form) of the transition amplitude. In the post form the bound state electron wave function $\phi$ is assumed to be free from the interaction for $t\rightarrow \infty$ while the free (i.e. unbound) positron wave function $\Psi$ feels both the laser and the Coulomb field. The laser four potential $A_L^{\mu}$ is considered as the interaction in this case and is turned off asymptotically at $t\rightarrow \infty$.

It should be noted that pair production is very similar to strong-field ionization \cite{Joachain00,Becker02,Milosevic03} from a theoretical point of view. In ionization an electron is lifted from a bound state to a state in the positive-energy continuum via the absorption of photons. Similarly, in pair production an electron in the negative energy continuum is lifted into a bound state. Hence, initial and final states are essentially interchanged. While for pair production we describe the process in the post form, in ionization it is more appropriate to use the prior form \cite{Reiss90,Crawford98}.

To be able to advance analytically one applies to Eq. (\ref{s-matrix}) the so called \textit{Strong-Field-Approximation}  (SFA, \cite{Keldysh65,Faisal73,Reiss80b,Reiss90b,Reiss92}), where one replaces the fully interacting positron wave function by a Volkov solution. The SFA can be applied when the laser field is sufficiently strong and the influence of the Coulomb potential on the positron is negligible. On the other hand the laser field strength must not be too strong so that the existence of bound states is assured. This sets limits on the nuclear charge number $Z$ which should be of intermediate value to give reasonable results \cite{Muller03b}.

Within the SFA, the amplitude in Eq. (\ref{s-matrix}) becomes
\begin{equation} \label{SFA}
S_{\mathrm{post}}^{\mathrm{SFA}} = - i e \int d^4x \overline{\phi} \slashed{A}_L \Psi^{\left( +\right)  }_{\mathrm{Volkov}} \,\,.
\end{equation} 

The so called \textit{Volkov} states \cite{Volkov35,Landau91} are solutions of the Dirac equation for a positron in a plane-wave field. They are defined as:

\begin{equation}
\Psi^{\left( +\right) }_{\mathrm{Volkov}}\left( x\right) = N_p \left( \mathds{1} + \frac{e \slashed{k}\slashed{A}_L}{2\left( k p\right) }\right) v_{p,s} \exp\left\lbrace i \mathcal S^{\left( +\right) } \right\rbrace \,\,,
\end{equation} 

where $p^{\mu}$ and $k^{\mu}$ are the four momenta of positron and laser photons, $v_{p,s}$ are the conventional Dirac spinors (chosen like in \cite{Bjorken64}) and the action reads

\begin{equation} \label{action}
\mathcal S^{\left( +\right) } = \left( p x\right) + \frac{e}{\left( k p\right) } \int^{\eta} \left[ p \cdot A\left( \tilde{\eta}\right) - \frac{e}{2} A^2\left( \tilde{\eta}\right)\right] d\tilde{\eta} \,\,.
\end{equation} 

The normalization reads $N_p = \sqrt{\frac{m}{q^0}}$ where $q^{\mu}=p^{\mu} + \frac{e^2 a^2}{4 \left( k^{\mu} p\right)} k$ is the effective four momentum in the presence of the laser field \cite{Landau91}.

In the present work, a linear laser polarization is chosen. The electric field points in the x-direction while the laser field propagates in the z-direction. Furthermore, temporal gauge is applied:

\begin{equation*}
A^{\mu}_L = a \left( 0,1,0,0\right) \cos\eta \,\,,\hspace{1cm} k^{\mu} = \omega\left( 1,0,0,1\right). \end{equation*}

Here, $a$ is the amplitude of the laser field, $\omega$ its frequency and $\eta = \left(  k x\right)$ the laser phase. The action in  Eq.(\ref{action}) can now be written explicitly as


\begin{equation} \label{action2}
\mathcal S^{\left( +\right) } = \left( q x\right) - \frac{e a}{\left( k p\right) } p_x \sin\eta + \frac{e^2 a^2}{8 \left( k p\right)} \sin 2\eta \,\,.
\end{equation} 

$\mathcal S^{\left( +\right) }$ enters as a phase factor in Eq. (\ref{SFA}). It can be expanded into a Fourier series, yielding the \textit{generalized} Bessel functions of two arguments \cite{Reiss90}:

\begin{equation} \label{genbessel}
 \tilde{J}_n\left( \alpha_1,\alpha_2 \right) = \sum_{m=-\infty}^{\infty}J_{n-2m}\left( \alpha_1 \right) J_m\left( \alpha_2 \right)
\end{equation}
The $J_n$ denote the regular Bessel functions of first kind. Their arguments $\alpha_1=\frac{e a }{\left( k p\right) } p_x$ and $\alpha_2 =\dfrac{-e^2 a^2}{8 \left( k p\right) }$ are the prefactors in Eq. (\ref{action2}).

\subsection{Ground-state}
The calculation is performed in the ion rest frame, where the bound-state wave function in the ground state takes the form:

\begin{equation}
\phi_{\mathrm{1s}}^s = g\left( r\right) \chi^s e^{-i E_{1s} t}                                         
\end{equation} 

It separates in a radial part $g\left( r\right)=C \left( 2 Z r / a_B \right) ^{\sigma-1} e^{-Z r /a_B}$, a spinor part $\chi^s$ and the time evolution.
$C$ is a normalization constant, $Z$ the nuclear charge number, $a_B=1/\left(\alpha m\right)$ the Bohr radius  and $\sigma=\sqrt{1-\alpha Z}$ the energy in units of the electron mass, i.e. $E_{1s}=m\sigma$ is the (relativistic) bound energy for the 1s state.
The explicit forms of the spinor $\chi^s$ and $C$ can be found in \cite{Eichler95,Bjorken64}.

In order to evaluate the pair production probability we follow the procedure outlined in \cite{Reiss90} for relativistic strong-field ionization of K-shell electrons (see also \cite{Muller03b}). First we sum over the possible spin configurations of the electron and positron, thus assuming a polarization-insensitive measurement of the produced particles. Furthermore we have to calculate the square of the S-matrix amplitude according to


\begin{equation}
 \mathcal{J} \equiv \sum_{s_+,s_-} \vert S_{\mathrm{post}}^{\mathrm{SFA}}\vert^2 \,\,.
\end{equation} 

The sum over the electron spin is directly performed by adding the two spinor pairs $M = \sum_{s_-} \chi_{s_-} \overline{\chi}_{s_-}' = \chi_{+1/2} \overline{\chi}_{+1/2}' + \chi_{-1/2} \overline{\chi}_{-1/2}'$ whereas for the positron we proceed by expressing the matrix elements of $M$ in terms of $\gamma$ matrices, applying the well-known theorem $\sum_{s_+} v_{\beta} \overline{v}_{\alpha} = \frac{1}{2m} \left(\slashed{p}-m\right)_{\beta \alpha}$ and calculating the traces via trace technology \cite{Bjorken64}. 
The time integral leads to the energy conserving $\delta$-function and the spatial integrals can be performed by elementary methods.

The calculation naturally splits into three parts denoted by the subscript letters $A$,$B$,$C$. The final result reads:

\begin{widetext}
\begin{equation} \label{finalresult}
\mathcal{J}=\frac{8  \pi^2}{E_q \left( {Z}/{a_B}\right) ^3} T \sum_{n \geq n_0} \frac{\delta\left( E_{1s}+E_q-n\omega\right) }{\left[ 1+\left( \rho  a_B /Z\right)^2 \right] ^4} \left[ u_A + u_B + u_C \right] 
\end{equation} 

$E_q=q^0$ is the dressed positron energy, $T = 2\pi \delta\left( 0 \right)$ the interaction time \cite{Bjorken64} and $n_0$ the minimum photon number due to energy conservation. The $\delta$-function shows the energy conservation  requirement explicitly. The term in the denominator represents the square of the Fourier transform of the (nonrelativistic) hydrogenlike wave function $\phi$. It plays an important role in the spatial distribution of the emitted positrons (see Sec. \ref{Results}). The functions $u_A$,$u_B$ and $u_C$ are defined as follows:

\begin{equation*}
\begin{split}
 u_A &=  e^2  a^2 \mathcal{P} \left(\tilde{J}_{n-1}^2+2\tilde{J}_{n-1}\tilde{J}_{n+1}+\tilde{J}_{n+1}^2\right)  \Big[ \left( p_0 + m \right) \sigma^2 \left( \frac{\rho a_B}{Z} \right)^4 \mathcal{U}^2 
\\ 
&\hspace{5mm} + \left( p_0 - m \right) \tau^2 \left( \frac{\rho a_B}{Z} \right)^2 \mathcal{V}^2 
+ 2  \sigma \tau \frac{a_B}{Z} \left( \frac{\rho a_B}{Z} \right)^2 \mathcal{U} \mathcal{V} \left( -p_x^2 + p_y^2  + p_z \left( p_z - b k_z \right) \right) \Big]
\end{split}
\end{equation*}

\begin{equation*}
 \begin{split}
u_B &= e a \nu \omega  \mathcal{P} 
\left(\tilde{J}_{n-2}\tilde{J}_{n-1}+\tilde{J}_{n-2}\tilde{J}_{n+1}+2\tilde{J}_{n}\tilde{J}_{n-1}+2\tilde{J}_{n}\tilde{J}_{n+1}+\tilde{J}_{n+2}\tilde{J}_{n-1}+\tilde{J}_{n+2}\tilde{J}_{n+1} \right) \\
&\hspace{5mm} \times \Big[ - p_x \sigma^2 \left( \frac{\rho a_B}{Z} \right)^4 \mathcal{U}^2 - p_x \tau^2 \left( \frac{\rho a_B}{Z} \right)^2 \mathcal{V}^2 + 2 p_x\left( p^0-2 p_z + b k_z \right) \sigma \tau \frac{a_B}{Z} \left( \frac{\rho a_B}{Z} \right)^2 \mathcal{U} \mathcal{V} 
\Big] \\
 \end{split}
\end{equation*}

\begin{equation*}
 \begin{split}
u_C &= \frac{1}{2} \nu^2 \omega^2 \mathcal{P} \left(p^0-p_z\right) 
 \left[ \sigma^2 \left( \frac{\rho a_B}{Z} \right)^4 \mathcal{U}^2  + \tau^2 \left( \frac{\rho a_B}{Z} \right)^2 \mathcal{V}^2  + 2\left( p_z - b k_z \right) \sigma  \tau \frac{a_B}{Z} \left( \frac{\rho a_B}{Z} \right)^2 \mathcal{U} \mathcal{V} \right] \\
&\hspace{5mm} \times \left[\tilde{J}_{n-2}\tilde{J}_{n-2} + 4\tilde{J}_{n-2}\tilde{J}_{n} + 2\tilde{J}_{n-2}\tilde{J}_{n+2} + 4\tilde{J}_{n}\tilde{J}_{n} + 4\tilde{J}_{n}\tilde{J}_{n+2} + \tilde{J}_{n+2}\tilde{J}_{n+2} \right] 
 \end{split}
\end{equation*} 
\end{widetext}

The arguments $\alpha_1$ and $\alpha_2$ of the generalized Bessel-functions $\tilde J_n$ are suppressed here in order to maintain a better readability.
$\mathcal{U}$,$\mathcal{V}$,$\mathcal{P}$ are functions containing momenta, bound state energy and normalization:


\begin{equation} \label{UVX}
\begin{split}
\mathcal{U}&= \sin \mathcal{X} + \frac{a_B \rho}{Z} \cos \mathcal{X} \\
\mathcal{V}&= - \sigma \frac{a_B \rho}{Z} \cos \mathcal{X} + \left[ 1+\left( 1+\sigma\right) \left( \frac{a_B \rho}{Z}\right) ^2\right] \sin \mathcal{X} \\
\mathcal{X}&= \sigma \arctan \left( \frac{a_B \rho}{Z}\right) 
\\
\mathcal{P}&= \frac{\left(1+ \sigma \right) \left(\Gamma \left( \sigma\right) \right) ^2 2^{2\left(\sigma -1 \right) }}{\Gamma\left(1+2\sigma \right) }\frac{\left[1+\left( \frac{a_B \rho}{Z}\right) ^2\right]^{2-\sigma} }{\left( \frac{a_B \rho}{Z}\right) ^6}
\end{split}
\end{equation}

Here, $\vec{\rho}$  $= \vec{q} - n \vec{k} = \vec{p} - b \vec{k}$. Furthermore, $\tau = \frac{1-\sigma}{\alpha Z}$ and $\nu = \frac{e^2 a^2}{2\left( k p \right)}$.
%
%

For the present process of pair creation, we are interested in the production rate of electrons and positrons.
Therefore, we have to integrate over the momentum of the free positron which has to be done numerically. It is advantageous to consider the effective momentum $q^{\mu}=p^{\mu}+\frac{e^2 a^2}{4 \left( k p\right)} k^{\mu} $ which includes the effect of the ponderomotive motion of the positron. For small laser intensity the parameters $p$ and $q$ are approximately equal. The fully differential production rate reads

\begin{equation} \label{d3R}
d^3R = \frac{\mathcal{J}}{T} \frac{d^3q}{\left( 2 \pi \right) ^3} \,\,.
\end{equation}

Numerical results of Eq. (\ref{d3R}) are shown in Sec. \ref{Results}.

\subsection{$\mathbf{2s_{1/2}}$ and $\mathbf{2p_{1/2}}$ states}

In this subsection the capture into the $2s_{1/2}$- and $2p_{1/2}$-states is considered. The former gives the highest contribution to the pair production rate after the ground state, whereas the latter is the first state deviating from the spherical shape.

The wave function for the $2s$-state reads \cite{Eichler95}
\begin{equation} \label{wave2s}
 \phi_{2s}^s = g\left( r \right) \, \chi_{2s}^s\left( r \right) \, e^{-i E_{2s}t} \,\,.
\end{equation} 
As before,  $g\left( r \right) = N_{2s} r^{\sigma-1} e^{-\zeta r}$ contains the normalization and the radial part. Note that for the excited states also the spinor $\chi_{2s}^s$ depends on $r$. The new abbreviation $\zeta = m \sqrt{\frac{1-\sigma}{2}}$ is used here. The bound energy reads $E_{2s}=m \sqrt{\frac{1+\sigma}{2}}$.
The wave function of the $2p_{1/2}$-state quantized along the beam axis is of the same form as the $2s$-state with differences in the normalization and the spinor part \cite{Eichler95}. The two states are degenerate in the Dirac theory, i.e. $E_{2s}=E_{2p_{1/2}}$.

The general procedure for the calculation of the $2s$- and $2p_{1/2}$-states is as before, but since the wave functions are more complicated, the final result for the production rate is more involved. We obtain
\begin{widetext}

\begin{equation} \label{ergebnis_2s}
\frac{d^3R}{d^3q}=\frac{1}{\pi E_q \zeta^3} \sum_{n \geq n_0} \frac{\delta\left( E_{2s}+E_q-n\omega\right) }{\left[ 1+\left( \rho  / \zeta\right)^2 \right] ^4} \left[ u_A + u_B + u_C \right] \,\,.
\end{equation}

The general form of the pair production rate (\ref{ergebnis_2s}) is as in Eq. (\ref{finalresult}). Once again, the energy conserving $\delta$-function appears and the sum over all photon orders $n$ is to be performed. Also, the square of the Fourier transform of the radial part of the Schr\"odinger wave functions appears in the denominator. The coefficients for the capture into the $2s$-state read:


\begin{equation*} \label{u_A_2s}
 \begin{split}
u_A &= e^2  a^2 \mathcal{P}^{\left( 2s \right)} \left( \frac{\rho}{\zeta}\right)^2 \left(\tilde{J}_{n-1}^2+2\tilde{J}_{n-1}\tilde{J}_{n+1}+\tilde{J}_{n+1}^2\right) \\
&\hspace{5mm} \times \Bigg\lbrace \left( m + p^0 \right) \sigma^2 \left( \frac{\rho}{\zeta}\right)^2 \mathcal{W}_1 - \left( m - p^0 \right) \aleph^2 \mathcal{W}_2  
+ \left( -p_x^2+p_y^2+p_z\left( p_z-b k_z \right) \right)2 \aleph  \sigma \zeta^{-1} \mathcal{W}_3 \Bigg\rbrace \,\,,
\end{split}
\end{equation*} 

\begin{equation*} \label{u_B_2s}
 \begin{split}
  u_B &= e a \nu \omega \mathcal{P}^{\left( 2s \right)} \left( \frac{\rho}{\zeta}\right)^2  \Bigg\lbrace \left( p^0-p_z \right) 2 \aleph \sigma \zeta^{-1} \mathcal{W}_3 p_x 
 -p_x\left[ \sigma^2 \left( \frac{\rho}{\zeta}\right)^2 \mathcal{W}_1 + \aleph^2 \mathcal{W}_2 + 2 \aleph \sigma \zeta^{-1} \mathcal{W}_3 \left( p_z - b k_z \right)\right] \Bigg \rbrace \\
&\hspace{5mm} \times \left(\tilde{J}_{n-2}\tilde{J}_{n-1}+\tilde{J}_{n-2}\tilde{J}_{n+1}+2\tilde{J}_{n}\tilde{J}_{n-1}+2\tilde{J}_{n}\tilde{J}_{n+1}+\tilde{J}_{n+2}\tilde{J}_{n-1}+\tilde{J}_{n+2}\tilde{J}_{n+1} \right) 
%
%
\end{split}
\end{equation*} 
and
\begin{equation*} \label{u_C_2s}
 \begin{split}
  u_C &= \frac{1}{2} \nu^2 \omega^2  \left(p^0-p_z\right) \mathcal{P}^{\left( 2s \right)} \left( \frac{\rho}{\zeta}\right)^2  
\left\lbrace \sigma^2 \left( \frac{\rho}{\zeta}\right)^2 \mathcal{W}_1 + \aleph^2 \mathcal{W}_2 + 2 \aleph \sigma \zeta^{-1} \mathcal{W}_3 \left( p_z - b k_z \right) \right\rbrace 
 \\
&\hspace{5mm} \times \left[\tilde{J}_{n-2}\tilde{J}_{n-2} + 4\tilde{J}_{n-2}\tilde{J}_{n} + 2\tilde{J}_{n-2}\tilde{J}_{n+2} + 4\tilde{J}_{n}\tilde{J}_{n} + 4\tilde{J}_{n}\tilde{J}_{n+2} + \tilde{J}_{n+2}\tilde{J}_{n+2} \right] \,\,.
\end{split}
\end{equation*} 


Due to the energy degeneracy the final result, Eq. (\ref{ergebnis_2s}), holds also for the $2p_{1/2}$-state with the coefficients given by 

\begin{equation*} \label{u_A_2p}
 \begin{split}
u_A &= e^2  a^2 \mathcal{P}^{\left( 2p \right)} \left( \frac{\rho}{\zeta}\right)^2 \left(\tilde{J}_{n-1}^2+2\tilde{J}_{n-1}\tilde{J}_{n+1}+\tilde{J}_{n+1}^2\right)   \\
&\hspace{5mm}
\times \Bigg\lbrace \left( m + p^0 \right) \mathcal{W}_2 - \left( m - p^0 \right) \aleph^2 \sigma^2 \left( \frac{\rho}{\zeta}\right)^2 \mathcal{W}_1 
%
 -2 \aleph \sigma \zeta^{-1} \mathcal{W}_3 \left[ -p_x^2 + p_y^2 + p_z\left( p_z - b k_z \right) \right] \Bigg\rbrace \,\,,
\end{split}
\end{equation*} 

\begin{equation*} \label{u_B_2p}
 \begin{split}
  u_B &= e a \nu \omega \mathcal{P}^{\left( 2p \right)} \left( \frac{\rho}{\zeta}\right)^2 \Bigg\lbrace -\left( p^0 - p_z \right) 2 \aleph \sigma \zeta^{-1} p_x \mathcal{W}_3 
%
 - p_x \left[   \aleph^2 \sigma^2 \left( \frac{\rho}{\zeta}\right)^2 \mathcal{W}_1 + \mathcal{W}_2 - 2 \aleph \sigma \zeta^{-1} \left( p_z - b k_z \right) \mathcal{W}_3\right] 
\Bigg\rbrace \\
&\hspace{5mm} \times \left(\tilde{J}_{n-2}\tilde{J}_{n-1}+\tilde{J}_{n-2}\tilde{J}_{n+1}+2\tilde{J}_{n}\tilde{J}_{n-1}+2\tilde{J}_{n}\tilde{J}_{n+1}+\tilde{J}_{n+2}\tilde{J}_{n-1}+\tilde{J}_{n+2}\tilde{J}_{n+1} \right) \,\,,
\end{split}
\end{equation*} 
 and
\begin{equation*} \label{u_C_2p}
 \begin{split}
  u_C &= \frac{1}{2} \nu^2 \omega^2  \left(p^0-p_z\right) \mathcal{P}^{\left( 2p \right)} \left( \frac{\rho}{\zeta}\right)^2  
\left\lbrace \aleph^2 \sigma^2 \left( \frac{\rho}{\zeta}\right)^2 \mathcal{W}_1 + \mathcal{W}_2 - 2 \aleph \sigma \zeta^{-1} \left( p_z - b k_z \right) \mathcal{W}_3 \right\rbrace \\ 
&\hspace{5mm} \times \left[\tilde{J}_{n-2}\tilde{J}_{n-2} + 4\tilde{J}_{n-2}\tilde{J}_{n} + 2\tilde{J}_{n-2}\tilde{J}_{n+2} + 4\tilde{J}_{n}\tilde{J}_{n} + 4\tilde{J}_{n}\tilde{J}_{n+2} + \tilde{J}_{n+2}\tilde{J}_{n+2} \right] \,\,.
\end{split}
\end{equation*} 

\end{widetext}
New abbreviations used here are $w=\left[1 + \left( \rho/\zeta\right) ^2\right] ^{-1}$, $E= \sqrt{\frac{1+\sigma}{2}}$, $\aleph = \sqrt{\frac{1-E}{1+E}}$, $\gimel_1 = 2E$, $\gimel_2 = -\frac{2 \zeta}{2E-1}$ and $\gimel_3 = -\frac{2 \zeta}{2E+1}$. The normalization factors are slightly different from before:

\begin{equation*}
\begin{split}
 \mathcal{P}^{\left( 2s \right)}
&= \frac{\left(\Gamma \left( \sigma\right) \right) ^2 2^{2\left(\sigma -2 \right) }}{\Gamma\left(1+2\sigma \right) } w^{\sigma-2} \left( \frac{\rho}{\zeta}\right) ^{-6} \frac{\left( 1+E\right) \left(2E-1 \right) }{E}
\\
\mathcal{P}^{\left( 2p \right)}&=  \mathcal{P}^{\left( 2s \right)} \frac{2E+1 }{2E-1} \,\,.
\end{split}
\end{equation*}


Furthermore,
%

\begin{equation} \label{Ws}
 \begin{split}
\mathcal{W}_1 &= \left( \gimel_1 \mathcal U + \gimel_2 \zeta^{-1} \left( \sigma+1 \right) {_r}\mathcal U  \right)^2 \\
&\\
\mathcal{W}_2 &= \left( \left( \gimel_1 + 2 \right) \mathcal{V} + \gimel_2 \zeta^{-1} {_r}\mathcal{V} \right)^2 \\
&\\
\mathcal{W}_3 &=  \left( \gimel_1 + 2 \right) \gimel_1 \mathcal{U} \mathcal{V} + \left( \gimel_1 + 2 \right) \gimel_2 \zeta^{-1} \left( \sigma + 1 \right) {_r}\mathcal{U} \mathcal{V} \\
&\hspace{5mm} + \gimel_1 \gimel_2 \zeta^{-1} \mathcal{U} {_r}\mathcal{V} + \gimel_2^2 \zeta^{-2} \left( \sigma + 1 \right) {_r}\mathcal{U} {_r}\mathcal{V} \\
\mathcal{W}_4 &= \left( \gimel_1 \mathcal U + \gimel_3 \zeta^{-1} \left( \sigma+1 \right) {_r}\mathcal U  \right)^2 \\
&\\
\mathcal{W}_5 &= \left( \left( \gimel_1 - 2 \right) \mathcal{V} + \gimel_3 \zeta^{-1} {_r}\mathcal{V} \right)^2 \\
&\\
\mathcal{W}_6 &= \left( \gimel_1 -2  \right) \gimel_1  \mathcal{U}\mathcal{V} + \left( \gimel_1 -2  \right) \gimel_3 \zeta^{-1} \left( \sigma +1 \right)  {_r}\mathcal{U}\mathcal{V} \\
&\hspace{10mm} + \gimel_1 \gimel_3  \zeta^{-1} \mathcal{U} {_r}\mathcal{V} + \gimel_3^2 \zeta^{-2} \left( \sigma +1 \right)  {_r}\mathcal{U} {_r}\mathcal{V} \,\,.
\end{split}
\end{equation} 

The abbreviations $\mathcal U$,$\mathcal V$ and $\mathcal X$ are defined as before (see Eq. (\ref{UVX})), additionally

\begin{equation*}
 _r\mathcal{U} = \left( 1 - \left( \frac{\rho }{\zeta} \right)^2 \right) w \sin \mathcal{X} + 2 \frac{\rho}{\zeta} w \cos \mathcal{X}
\end{equation*} 
and
\begin{equation*}
\begin{split}
 _r\mathcal{V} &= \sigma\left( -2\left( \sigma + 1\right) w + 3 +2\sigma\right) \sin \mathcal{X} \\
&\hspace{3mm} - \sigma \left( 2\left( \sigma + 1\right) \left( \frac{\rho}{\zeta}\right)  w - \left( \sigma + 2 \right) \left( \frac{\rho}{\zeta}\right)\right) \cos \mathcal{X} \,\,.
\end{split}
\end{equation*}

Further analysis of the capture in the L-shell must be surveyed with numerical methods and will be discussed in the following section.

\section{Results}\label{Results}
In the following, results for the numerical integration of Eqs. (\ref{d3R}) and (\ref{ergebnis_2s}) are shown. Laser parameters of the planned X-ray free electron laser (XFEL) at DESY are assumed (see \cite{Desy02}). If not otherwise stated, the intensity parameter is $\xi=10^{-4}$, frequency $\omega=9$ keV, hence intensity $I = 7.2 \times 10^{17} \mathrm{W/cm^2}$. Furthermore, the ions have a nuclear charge number of $Z=50$ and are accelerated to a relativistic $\gamma$-factor of $\gamma=50$.
Primed quantities are in the ion rest frame whereas unprimed are in the laboratory frame. The total rate is not a Lorentz invariant since the relativistic time dilation has to be taken into account. The transformation law between the two frames simply reads
\begin{equation}
 R_{\mathrm{lab}}=\frac{1}{\gamma} R_{\mathrm{ion}}' \,\,.
\end{equation} 
In the multiphoton regime the process of lowest photon order will dominate - for the parameters introduced above the minimal photon number is $n_0=2$. Consequently, we will mainly discuss the two-photon process.

The created electrons can be bound in the ground state but generally also in all higher states. Since the momentum spread is substantially larger for the ground state electrons, the capture probability to the K-shell is expected to be the highest \cite{Eichler95}. Capture into the L-shell yields the main correction to the total pair production rate. Hence, up to Sec. \ref{highint} we will focus on the capture in the K-shell and work out the influence of the L-shell in Sec. \ref{l-shell}.

\subsection{One-Photon limit}

Pair production via the absorption of single $\gamma$-photons has been considered before (see, e.g., \cite{Sauter31,Agger97}). The main goal of the present work is to regard nonlinear processes, thus exploiting the coherence of laser light. However, it is interesting to compare results obtained by the SFA methods of this work with earlier results of single photon calculations, where the radiation field is treated perturbatively.

\begin{figure}
 \centering
 \includegraphics[width=1.0 \linewidth,bb=50 50 266 201]{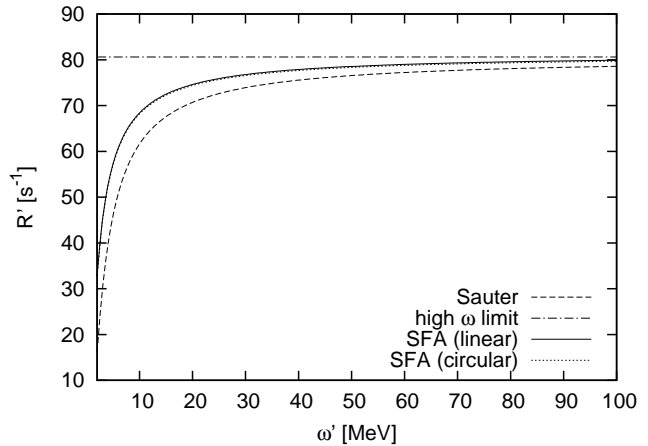}
 \caption{Comparison of different calculations for the frequency dependence of the total pair production rate in the case of the absorption of a single photon (here, $\xi=10^{-4}$, $Z=1$, $\gamma=50$).}
 \label{sauter}
\end{figure}

Figure \ref{sauter} shows the scaling of the pair-production rate with the frequency of the laser for the absorption of a single high-energy photon (here, $Z=1$). For this linear process, i.e. the absorption of just one photon, several calculations are compared. 
A calculation in Born-approximation by Sauter \cite{Sauter31,Landau91} and its high frequency limit are shown. We compare the frequency dependency of the total rate to SFA-calculations for linear and circular polarization. One can see that the two polarizations yield the same rates as is expected for the absorption of a single photon.
The main difference between the SFA calculation and the calculation in Born-approximation is that in the first case the positron is described by Volkov wave functions whereas in the latter the positron is a free wave. The two different calculation show a very similar frequency dependence and obtain the same high-frequency limit.
Hence, the comparison of our calculations to previously obtained results supports the applied methods and approximations.

\subsection{Scaling of total rate for 2-photon absorption} \label{totalrate}
In this section, results for the scaling of the total rate for the absorption of two photons are shown.

\begin{figure}
 \centering
 \includegraphics[width=1.0\linewidth,bb=50 50 266 201]{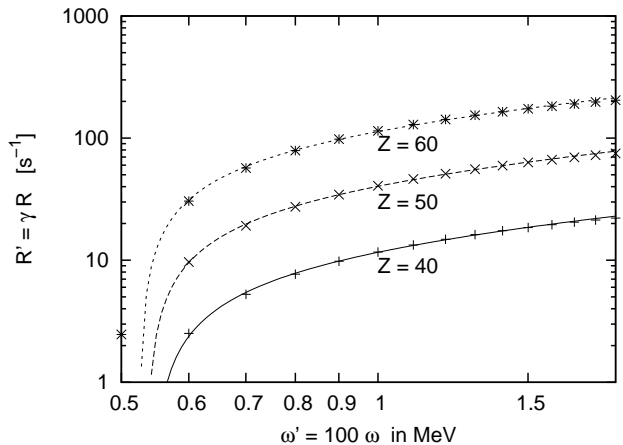}
 \caption{Dependence on frequency for various nuclear charge numbers $Z$. The lines are fits over calculated data points (here, $\xi=10^{-4}$, $\gamma=50$, $n=2$).}
 \label{omegaZ}
\end{figure}

Figure \ref{omegaZ} shows the scaling of the total rate with the laser frequency for different atomic charge numbers $Z$. For $\omega = 6 - 16$ keV the rate scales with $\left( \omega' - \omega_{\mathrm{min}} \right)^{1/2}$ for all $Z$, where $\omega_{\mathrm{min}}$ is the minimal frequency originating from the energy conservation requirement. Above, the rate approaches a constant value due to the saturation of the available phase space.
Below $6$ keV, i.e. close to the energetic threshold, the rate's dependence on the frequency can still be described by a power law but the power is slightly larger than $r=1/2$. 
It is interesting that the scaling-law exponent $r$ depends on the laser polarization. For a circularly polarized field the value $r=1$ was found \cite{Muller03b}.
A similar behaviour has been derived for free pair creation by two-photon absorption \cite{Milstein06}: In the case of linear (circular) polarization the cross section close to the energy threshold follows a power law with $r=2 \,\,\left( 4 \right)$. I.e., for both pair production channels, circular laser polarisation leads to an exponent which is twice as large as for linear polarisation.

\begin{figure}
 \centering
 \includegraphics[width=1.0 \linewidth,bb=50 50 266 201]{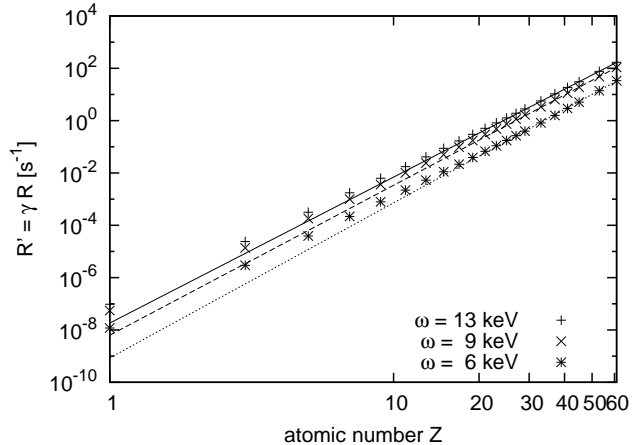}
 \caption{Dependence of the pair production rate on the atomic number $Z$ for various frequencies. The lines represent fits over calculated data points (here, $\hbar \omega = 6$ keV (dotted), $\hbar \omega = 9$ keV (dashed), $\hbar \omega = 13$ keV (solid), $\xi=10^{-4}$, $\gamma=50$).}
 \label{Zatom}
\end{figure}

Figure \ref{Zatom} shows the scaling of the total rate with the atomic charge number $Z$.
A power law $\sim Z^d$ describes the calculated values well, we find a power $d$ of $5.9$, $5.7$ and $5.6$ for photon energies of $6$, $9$ and $13$ keV, respectively. For capture processes one generally expects a power close to $5$ \cite{Eichler95}. The fitted curves in Figure \ref{Zatom} slightly deviate from the data points for small and high $Z$ values.
%
When we consider only small $Z$, i.e. only weak Coulomb fields (hence small deviation of the SFA), the power approaches the value $5$ in the high frequency limit. The same holds for the Born-approximation calculation by Sauter \cite{Sauter31}. Agger and S\o{}rensen found in a calculation for one photon pair production taking into account the full Coulomb effects a power slightly below $5$ \cite{Agger97}. 

In summary, the final scaling equation for bound-free pair production by the absorption of $2$ photons from a linearly polarized laser field reads: 

\begin{equation} \label{scaling}
 R \propto \gamma^{-1}\xi^{4} Z^{5.7} \left( \omega' - \omega_{\mathrm{min}} \right)^{1/2}
\end{equation} 

A similar scaling behavior exists for higher photon orders. The only difference is in the power of $\xi$ - for the absorption of $n$ photons it scales with $\xi^{2n}$. Since $\xi \ll 1$ in the multiphoton regime, the higher order terms are considerably smaller.

For free-free pair creation the rate grows with $Z^2$, hence the portion of bound-free pair creation becomes increasingly important for higher $Z$ values.


\subsection{Angular distributions in ion frame} \label{sec:angulardis}

\begin{figure}
 \includegraphics[width=1.0 \linewidth,bb=50 50 266 201]{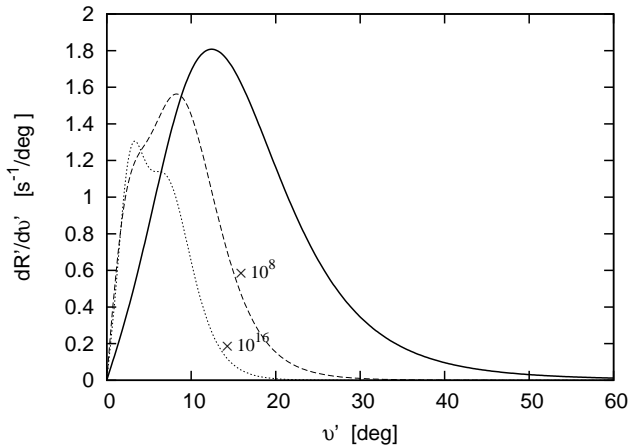}
 \caption{Polar angular distribution of created positron in the nuclear rest frame (here, $\hbar \omega =9$ keV, $\xi=10^{-4}$, $\gamma=50$, $Z=50$) for the absorption of two (solid line), three (dashed) and four (dotted) photons.}
 \label{theta}
\end{figure}

The free positron is emitted under a solid angle $d\Omega'=d\cos\vartheta'd\varphi'$. The distribution of the polar angle $\vartheta'$ is shown in Fig. \ref{theta} where $\vartheta'=0$ corresponds to the laser propagation direction. One can see that no positrons are emitted at $\vartheta = 0$. 
The distribution has a peak at a small angle and falls off quickly thereafter. 
The general form results from the term $\left[ 1+\left( \rho  a_B /Z\right)^2 \right] ^{-4}$ in the equation for the final analytic result (\ref{finalresult}). It represents the square of the Fourier transform of the (nonrelativistic) bound wave function.
Compared are the distributions for higher photon orders which have additional structure. The latter is a property of the generalized Bessel functions (\ref{genbessel}) and therefore a multiphoton effect. The higher photon order rates have been scaled by $10^{8\left( n-2 \right)}$ in order to be visible in this graph. The scaling of the rate with the intensity parameter is $\xi^{2n}$ (see Sec. \ref{totalrate}).

\begin{figure}
 \includegraphics[width=1.0 \linewidth,bb=50 50 266 201]{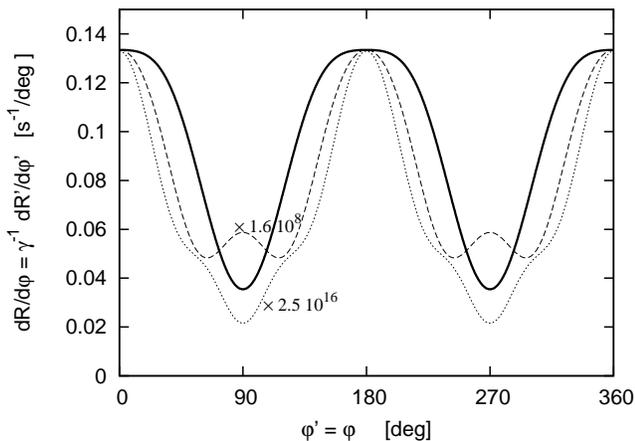}
 \caption{Azimuth angular distribution of the created positron in the nuclear rest frame (here, $\hbar \omega =9$ keV, $\xi=10^{-4}$, $\gamma=50$, $Z=50$) for the absorption of two (solid line), three (dashed) and four (dotted) photons.}
 \label{phi}
\end{figure}

For the case of a linearly polarized laser a dependence on the azimuth angle is expected (see Fig. \ref{phi}). The emission is maximal in direction of the electric field component of the laser field and minimal but non-zero in direction of the magnetic field component. For higher photon orders some additionally structure is present so that for odd photon numbers the minimum is shifted. 
The azimuthal distribution is the same (up to total scaling by $1/\gamma$) in the lab frame since the azimuth angle is not affected by the Lorentz transformation.

The non-isotropic azimuthal dependence of the pair production rate is clearly a polarization effect and absent in the case of circular laser polarization considered before \cite{Muller03b}. The positron distributions in the polar angle, however, are similar for both polarization states. Regarding the total creation rates,
we find that at the same intensity parameter $\xi$ the circular case is favored, but when considering the same intensity ($I = 7.2 \times 10^{17} \mathrm{W/cm^2}$), the rate for linear polarization ($R_{\mathrm{lin}}'= 34.4 \,\mathrm s^{-1}$) is slightly larger than for circular polarization ($R_{\mathrm{circ}}'= 26.0 \,\mathrm s^{-1}$). This comes from the fact that at same intensity the peak electric field of a linear polarized wave is higher (by a factor of $\sqrt{2}$).

%
%

\subsection{Transformation to the laboratory frame}

\begin{figure}
 \centering
 \includegraphics[width=1.0 \linewidth,bb=50 50 266 201]{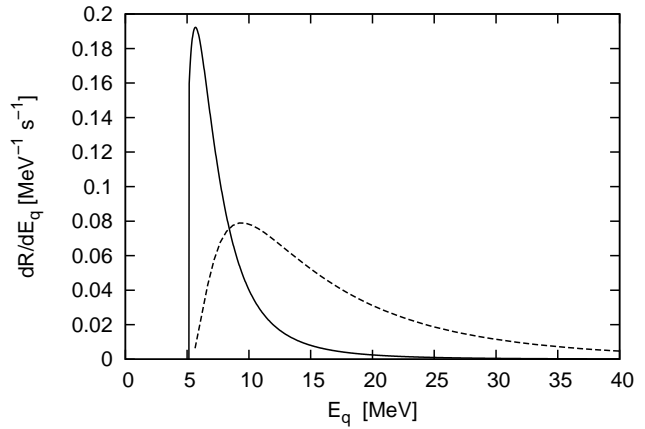}
 \caption{Rate dependence on the energy in the laboratory frame for the bound-free (solid line) and free-free (dashed) channel (here, $\hbar \omega = 9$ keV, $\xi=10^{-4}$, $\gamma=50$, $Z=50$, $n=2$).}
 \label{e_lab}
\end{figure}

For experimental observation it is essential to consider the process in the laboratory frame. When one considers the polar angular rate or energy distribution the transformation is no longer as simple as for the total rate. Instead Lorentz transformations have to be performed \cite{Eichler95} and unlike before the energy of the emitted positrons is no longer fixed. 
In the ion frame, energy conservation fixes the positron energy to $E_q'=1.33$ MeV.
Now, for each angle $\vartheta'$ in the ion frame, there exists an energy $E_q$ in the lab frame. Figure \ref{e_lab} depicts the energy-differential rate. It exists a minimal and a maximal energy corresponding to $\vartheta' = 0 $ and $ \pi$ and the distribution is peaked at $E_q = 6.57$ MeV which matches the maximum of the $\vartheta'$-distribution. The positron energies are very high, the positron's relativistic $\gamma$-factor is $\gamma_+ = \frac{E_q}{m} \gtrsim 10$ so that the emitted positrons are highly relativistic particles. The distribution for the free-free case is broader which is due to a broader angular distribution in the ion frame.

\begin{figure}
 \centering
 \includegraphics[width=1.0 \linewidth,bb=50 50 266 201]{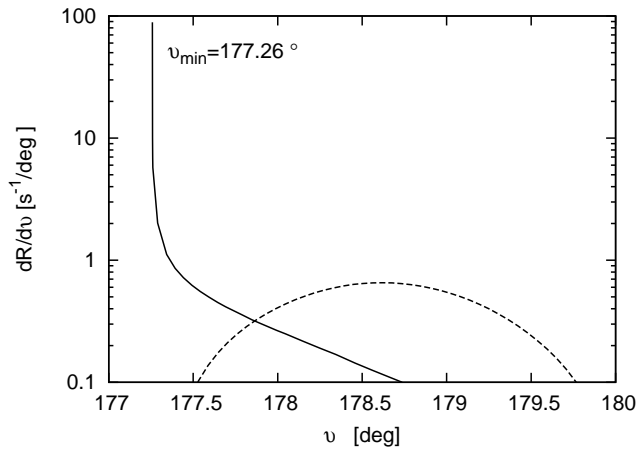}
 \caption{Rate dependence on the polar angle in the laboratory frame for the bound-free (solid line) and free-free (dashed) channel (here, $\hbar \omega =9$ keV, $\xi=10^{-4}$, $\gamma=50$, $Z=50$, $n=2$).}
 \label{theta_lab}
\end{figure}

Similarly, the polar angular distributions for bound-free and free-free pair production differ in the lab frame, as shown in Fig. \ref{theta_lab}. Basically all positrons are emitted under the same angle of $\vartheta_{\mathrm{min}} = 177.26^{\circ}$ in the case of the bound-free channel.
At smaller angles the emission of positrons is kinematically forbidden while at larger angles the emission is nearly completely suppressed. Pictorially, the angles in Fig. \ref{theta} are squeezed into one due to the high velocity of the ion.
This results from the form of the Jacobi-determinant that appears in the Lorentz transformation. It is proportional to $\left( 1-\beta/\beta_+' \cos \vartheta'\right)^{-1}$, where $\beta$ and $\beta_+'$ are the relativistic velocities of ion (in the lab frame) and positron (in the ion frame), respectively. Since $\beta > \beta_+'$, the determinant formally diverges at one angle $\vartheta_{\mathrm{min}}$, yielding the characteristic signature of the angular distribution. It should be noted however, that it is an integrable divergence not leading to difficulties in an experimental observation. The minimum angle $\vartheta_{\mathrm{min}}$ coincides with the minimal accessible angle in the lab frame, $\sin \vartheta_{\mathrm{min}} = \gamma_+ \beta_+ / \gamma \beta$, which is described in \cite{Eichler95}. For circular polarization the same angle $\vartheta_{\mathrm{min}}$ was found \cite{Muller03b}.
%
It is noteworthy that the angle of emission is unequal to $180^{\circ}$ which would have been difficult for a possible experiment. Compared in Fig. \ref{theta_lab} is the angular distribution for the competing free-free channel. The positron distribution is also confined, but by far less than in the bound-free case. In an experiment this allows to distinguish both channels by positron detection only. This is important because the bound state can decay via photoionization.
For the given parameters the two pair production processes have approximately the same total rate ($R_{\mathrm{free-free}} = 0.98 \, \mathrm{s}^{-1}$, $R_{\mathrm{bound-free}} = 0.69 \, \mathrm{s}^{-1}$) but scale differently with $Z$ and $\omega$.
To obtain an estimate of experimental pair production rates one should take into account the pulse length and repetition rate as well as the projectile beam density. For example, when $N\sim 10^{10}$ ions collide with a laser beam of $T\sim 100$ fs duration and $f\sim 1$ kHz repetition rate \cite{Desy02}, then the number of pair creation events per second is of the order of $R T N f /2 \sim 1 \mathrm s^{-1}$ (assuming perfect beam overlap).

\subsection{High intensity domain} \label{highint}

\begin{figure}
 \centering
 \includegraphics[width=1.0\linewidth,bb=50 50 266 201]{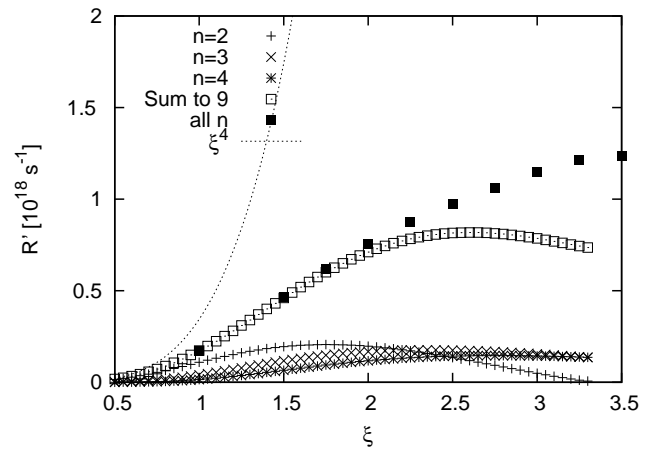}
 \caption{Bound-free Pair production rates in the $\xi \sim 1$ domain for different photon orders. Also shown is the sum of the simultaneous absorption of $9$ photons (open squares) and an estimate for the sum over all photon orders (solid squares) (here, $\hbar \omega =9$ keV, $\gamma=50$, $Z=50$).}
 \label{highXi}
\end{figure}

For a relatively low intensity parameter $\xi \ll 1$, the pair production rate scales with $\xi^{2n}$. The lowest possible photon order gives by far the largest contribution to the rate. 
If one increases the intensity, the higher photon orders become important whereas the two-photon process peters out at some intensity (see Fig. \ref{highXi}). The phase space for the two-photon process diminishes as one increases the intensity and at one point the process is no longer possible due to energy conservation: The ponderomotive energy of the positrons becomes too large. Eventually, the same happens successively for the higher photon orders but the black squares in Fig. \ref{highXi} indicate that the sum over all photon orders continues to increase. To estimate the contribution of all photons we calculated for each specific $\xi$ the rate for photon numbers from $2$ to $9$ and found out that they decrease exponentially for the larger $n$ values. From the exponential fit we were able to perform the sum to infinity, hence giving us an approximate result for all photon orders.
We point out that the effect described here is analogous to the phenomenon of channel closing in above-threshold ionization of atoms \cite{Kopold02}.

The results of this section are obtained by the same formalism as before; note, however, that the existence of bound states in the $\xi \sim 1$ domain is no longer evident since the laser field strength is very large and even exceeds the atomic binding field. Therefore, the results should be considered with some care.

\subsection{L-shell contribution} \label{l-shell}

\begin{figure}
 \centering
 \includegraphics[width=1.0 \linewidth,bb=50 50 266 201]{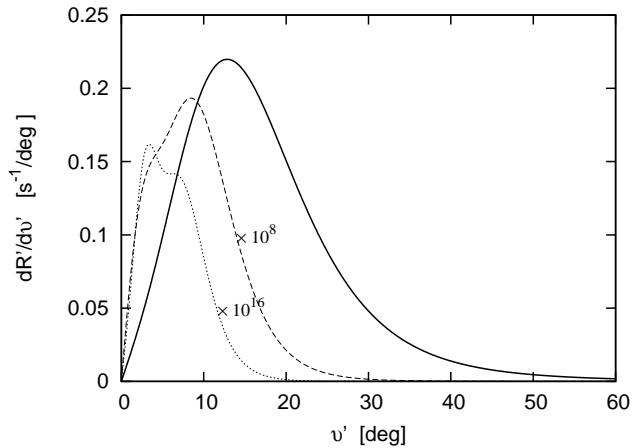}
 \caption{Polar angular distribution of the created positron for the capture in the $2s$-state in the nuclear rest frame (here, $\hbar \omega =9$ keV, $\xi=10^{-4}$, $\gamma=50$, $Z=50$) for the absorption of two (solid line), three (dashed) and four (dotted) photons.}
 \label{fig:theta_2s}
\end{figure}

Apart from the capture in the ground state, the electrons can also be bound in higher states.
Fig. \ref{fig:theta_2s} shows the distribution of the polar angle for different photon orders and the capture in the $2s$-state. It is very similar to the distribution for the capture in the $1s$-state (Fig. \ref{theta}). The total rate, however, is substantially smaller. Table \ref{ratiorates} shows absolute values of pair production rates for the $1s$- and $2s$-states for different nuclear charge numbers $Z$. Their ratio $2s$/$1s$ is nearly constant and approximately $0.125$. Furthermore, we find the same ratio for higher photon orders. Pratt \cite{Pratt60b} proved that for one-photon pair production in the high energy limit the ratio of the $ns$-shell to the K-shell is $\sigma(ns)=\sigma(K)/n^3$ (here, $n$ denotes the principal quantum number). Thus, we find that this also holds for higher photon numbers.

\begin{table}
\newcommand{\mc}[3]{\multicolumn{#1}{#2}{#3}}
\begin{tabular}{ll|l|l}
\mc{1}{c}{\bfseries 1s} & \mc{1}{c}{\bfseries 2s}  & \mc{1}{|c}{\bfseries 2s/1s}  & \mc{1}{|c}{Z}  \\ \hline
\mc{1}{c}{$5.53 \times 10^{-8}$} & \mc{1}{c}{$6.91\times 10^{-9}$}  & \mc{1}{|c}{$1.25\times 10^{-1}$}  & \mc{1}{|c}{1} \\ 
\mc{1}{c}{$3.44\times 10^{1}$} & \mc{1}{c}{$4.33\times 10^{0}$}  & \mc{1}{|c}{$1.26\times 10^{-1}$}  & \mc{1}{|c}{50}  \\ 
\mc{1}{c}{$5.24\times 10^{2}$} & \mc{1}{c}{$6.41\times 10^{1}$}  & \mc{1}{|c}{$1.22\times 10^{-1}$}  & \mc{1}{|c}{80}  \\ 
\end{tabular}
\caption{Total rates in the ion frame (in s$^{-1}$) for different states and nuclear charge numbers $Z$ (here, $\hbar \omega = 9$ keV, $\xi = 10^{-4}$, $\gamma =50$, $n=2$).}
\label{ratiorates}
\end{table}

The calculation of the $2p_{1/2}$-state is very similar to the $2s$-state. However, the total pair production rates are substantially lower. Unlike before, for the capture into the $2p_{1/2}$-state no simple scaling law for the ratio to the ground state can be found. 
But whereas the $2s$-state scales with the same power of the nuclear charge number $Z$ as the $1s$-state ($\sim Z^{5.7}$), this is no longer true for $2p_{1/2}$. Calculations for one-photon pair creation in Born-approximation suggest a power of $Z^7$ and we find $Z^{7.5}$. At $Z=1$ the total rate of the $2p_{1/2}$ is a negligible fraction to the $1s$ state ($\sim 10^{-4}$) but steadily increases to about $5\%$ at $Z=92$. Similar contributions were found by Agger et. al. for one-photon absorption \cite{Agger97}; the $2p_{3/2}$-state yields another $\sim 2\%$ at high nuclear charge numbers $Z$ in this case.

\begin{figure}
 \centering
 \includegraphics[width=1.0 \linewidth,bb=50 50 266 201]{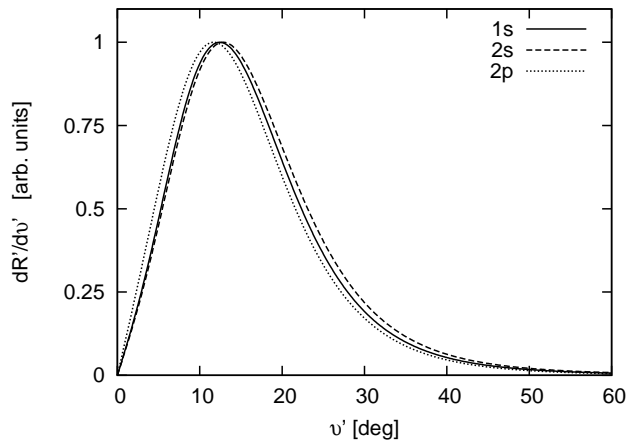}
 \caption{Comparison of the (scaled) polar angular distributions of the emitted positrons for the capture into different atomic states.}
 \label{theta_K_L}
\end{figure}

Fig. \ref{theta_K_L} shows the (scaled) polar angular distribution for the capture into the $1s$-, $2s$- and $2p_{1/2}$-states. One can see that the distributions are very much alike. At first sight this might appear surprising since we found in Sec. \ref{sec:angulardis} that the general structure is governed by the terms in Eqs. (\ref{finalresult}) and (\ref{ergebnis_2s}) which come from the square of the Fourier transform of the nonrelativistic bound state wave functions. When plotted as a function of $\rho'=\left( q'^2 + n^2\omega'^2 - 2 n  \omega' q' \cos \vartheta'\right)^{1/2}$ the width of the $1s$ state in momentum space is twice as large as the width of the $2s$- and $2p_{1/2}$-states. But when plotted as a function of $\vartheta'$, the width of the distributions turn out to coincide for all three states.

In conclusion, the L-shell gives an additional contribution to the total bound-free pair production rate of $15 - 20 \%$. The next higher state, the $3s$-state, can be estimated to increase the rate by $1/27\approx 3\%$, according to \cite{Pratt60b}.

\section{Conclusion} \label{Conclusion}

In this paper we considered nonlinear bound-free electron-positron pair production in the collision of a relativistic bare ion with an intense X-ray laser field. The scaling of the total rate and the angular distributions of the emitted positrons in the ion rest frame and the laboratory frame were analyzed.
For the first time, bound-free pair production with a many-cycle, linearly polarized laser beam was investigated. Distinct polarization effects were found in the azimuthal distribution and different frequency scaling.
Furthermore, we analyzed the capture of the electron into higher atomic shells and found that these give a contribution of $\sim 15 - 20 \%$ to the total bound-free pair production rate.
The bound-free channel is comparable in the total rate to the free-free pair production channel for high ionic charges. In a possible experiment one should be able to distinguish the two channels by observing the angular distribution of the emitted positrons.
For a laser like the planned XFEL at DESY the process could become observable when combining the laser with a sufficiently strong ion accelerator.
Apart from the fundamental significance of the process, a possible future application would be to use the described setup for measuring high laser intensities via the total rates \cite{Salamin06,Mourou06}. Moreover, if anti-protons were used as projectiles, the scheme might in principle be applicable for the creation of relativistic anti-hydrogen atoms \cite{Uggerhoj06}.


%
%

%

%



\begin{thebibliography}{64}
\expandafter\ifx\csname natexlab\endcsname\relax\def\natexlab#1{#1}\fi
\expandafter\ifx\csname bibnamefont\endcsname\relax
  \def\bibnamefont#1{#1}\fi
\expandafter\ifx\csname bibfnamefont\endcsname\relax
  \def\bibfnamefont#1{#1}\fi
\expandafter\ifx\csname citenamefont\endcsname\relax
  \def\citenamefont#1{#1}\fi
\expandafter\ifx\csname url\endcsname\relax
  \def\url#1{\texttt{#1}}\fi
\expandafter\ifx\csname urlprefix\endcsname\relax\def\urlprefix{URL }\fi
\providecommand{\bibinfo}[2]{#2}
\providecommand{\eprint}[2][]{\url{#2}}

\bibitem[{\citenamefont{Bertulani et~al.}(2005)\citenamefont{Bertulani, Klein,
  and Nystrand}}]{Bertulani05}
\bibinfo{author}{\bibfnamefont{C.~A.} \bibnamefont{Bertulani}},
  \bibinfo{author}{\bibfnamefont{S.~R.} \bibnamefont{Klein}}, \bibnamefont{and}
  \bibinfo{author}{\bibfnamefont{J.}~\bibnamefont{Nystrand}},
  \bibinfo{journal}{Ann. Rev. Nucl. Sci.} \textbf{\bibinfo{volume}{55}},
  \bibinfo{pages}{271} (\bibinfo{year}{2005}).

\bibitem[{\citenamefont{Baur et~al.}(2007)\citenamefont{Baur, Hencken, and
  Trautmann}}]{Baur07}
\bibinfo{author}{\bibfnamefont{G.}~\bibnamefont{Baur}},
  \bibinfo{author}{\bibfnamefont{K.}~\bibnamefont{Hencken}}, \bibnamefont{and}
  \bibinfo{author}{\bibfnamefont{D.}~\bibnamefont{Trautmann}},
  \bibinfo{journal}{Phys. Rep.} \textbf{\bibinfo{volume}{453}},
  \bibinfo{pages}{1} (\bibinfo{year}{2007}).

\bibitem[{\citenamefont{Eichler and Meyerhof}(1995)}]{Eichler95}
\bibinfo{author}{\bibfnamefont{J.}~\bibnamefont{Eichler}} \bibnamefont{and}
  \bibinfo{author}{\bibfnamefont{W.~E.} \bibnamefont{Meyerhof}},
  \emph{\bibinfo{title}{Relativistic atomic collisions}}
  (\bibinfo{publisher}{Academic Press}, \bibinfo{address}{London},
  \bibinfo{year}{1995}).

\bibitem[{\citenamefont{Brezin and Itzykson}(1970)}]{Brezin70}
\bibinfo{author}{\bibfnamefont{E.}~\bibnamefont{Brezin}} \bibnamefont{and}
  \bibinfo{author}{\bibfnamefont{C.}~\bibnamefont{Itzykson}},
  \bibinfo{journal}{Phys. Rev. D} \textbf{\bibinfo{volume}{2}},
  \bibinfo{pages}{1191} (\bibinfo{year}{1970}).

\bibitem[{\citenamefont{Salamin et~al.}(2006)\citenamefont{Salamin, Hu,
  Hatsagortsyan, and Keitel}}]{Salamin06}
\bibinfo{author}{\bibfnamefont{Y.~I.} \bibnamefont{Salamin}},
  \bibinfo{author}{\bibfnamefont{S.~X.} \bibnamefont{Hu}},
  \bibinfo{author}{\bibfnamefont{K.}~\bibnamefont{Hatsagortsyan}},
  \bibnamefont{and} \bibinfo{author}{\bibfnamefont{C.~H.}
  \bibnamefont{Keitel}}, \bibinfo{journal}{Phys. Rep.}
  \textbf{\bibinfo{volume}{427}}, \bibinfo{pages}{41} (\bibinfo{year}{2006}).

\bibitem[{\citenamefont{Maquet and Grobe}(2002)}]{Maquet02}
\bibinfo{author}{\bibfnamefont{A.}~\bibnamefont{Maquet}} \bibnamefont{and}
  \bibinfo{author}{\bibfnamefont{R.}~\bibnamefont{Grobe}}, \bibinfo{journal}{J.
  Mod. Opt.} \textbf{\bibinfo{volume}{49}}, \bibinfo{pages}{2001}
  (\bibinfo{year}{2002}).

\bibitem[{\citenamefont{Mourou et~al.}(2006)\citenamefont{Mourou, Tajima, and
  Bulanov}}]{Mourou06}
\bibinfo{author}{\bibfnamefont{G.}~\bibnamefont{Mourou}},
  \bibinfo{author}{\bibfnamefont{T.}~\bibnamefont{Tajima}}, \bibnamefont{and}
  \bibinfo{author}{\bibfnamefont{S.}~\bibnamefont{Bulanov}},
  \bibinfo{journal}{Rev. Mod. Phys.} \textbf{\bibinfo{volume}{78}},
  \bibinfo{pages}{309} (\bibinfo{year}{2006}).

\bibitem[{\citenamefont{Marklund and Shukla}(2006)}]{Marklund06}
\bibinfo{author}{\bibfnamefont{M.}~\bibnamefont{Marklund}} \bibnamefont{and}
  \bibinfo{author}{\bibfnamefont{P.}~\bibnamefont{Shukla}},
  \bibinfo{journal}{Rev. Mod. Phys.} \textbf{\bibinfo{volume}{78}},
  \bibinfo{pages}{591} (\bibinfo{year}{2006}).

\bibitem[{\citenamefont{Schwinger}(1951)}]{Schwinger51}
\bibinfo{author}{\bibfnamefont{J.}~\bibnamefont{Schwinger}},
  \bibinfo{journal}{Phys. Rev.} \textbf{\bibinfo{volume}{82}},
  \bibinfo{pages}{664} (\bibinfo{year}{1951}).

\bibitem[{\citenamefont{Materlik and Tschentscher}(2001)}]{Desy02}
\bibinfo{editor}{\bibfnamefont{G.}~\bibnamefont{Materlik}} \bibnamefont{and}
  \bibinfo{editor}{\bibfnamefont{T.}~\bibnamefont{Tschentscher}}, eds.,
  \emph{\bibinfo{title}{TESLA XFEL, Technical Design Report, Part V: The X-Ray
  Free Electron Laser}} (\bibinfo{address}{Hamburg}, \bibinfo{year}{2001}),
  \urlprefix\url{http://www.xfel.net}.

\bibitem[{\citenamefont{Burke et~al.}(1997)\citenamefont{Burke, Field,
  Horton-Smith, Spencer, Walz, Berridge, Bugg, Shmakov, Weidemann, Bula
  et~al.}}]{Burke97}
\bibinfo{author}{\bibfnamefont{D.~L.} \bibnamefont{Burke}},
  \bibinfo{author}{\bibfnamefont{R.~C.} \bibnamefont{Field}},
  \bibinfo{author}{\bibfnamefont{G.}~\bibnamefont{Horton-Smith}},
  \bibinfo{author}{\bibfnamefont{J.~E.} \bibnamefont{Spencer}},
  \bibinfo{author}{\bibfnamefont{D.}~\bibnamefont{Walz}},
  \bibinfo{author}{\bibfnamefont{S.~C.} \bibnamefont{Berridge}},
  \bibinfo{author}{\bibfnamefont{W.~M.} \bibnamefont{Bugg}},
  \bibinfo{author}{\bibfnamefont{K.}~\bibnamefont{Shmakov}},
  \bibinfo{author}{\bibfnamefont{A.~W.} \bibnamefont{Weidemann}},
  \bibinfo{author}{\bibfnamefont{C.}~\bibnamefont{Bula}}, \bibnamefont{et~al.},
  \bibinfo{journal}{Phys. Rev. Lett.} \textbf{\bibinfo{volume}{79}},
  \bibinfo{pages}{1626} (\bibinfo{year}{1997}).

\bibitem[{\citenamefont{Reiss}(1962)}]{Reiss62}
\bibinfo{author}{\bibfnamefont{H.~R.} \bibnamefont{Reiss}},
  \bibinfo{journal}{J. Math. Phys.} \textbf{\bibinfo{volume}{3}},
  \bibinfo{pages}{59} (\bibinfo{year}{1962}).

\bibitem[{\citenamefont{Ritus}(1972)}]{Ritus72}
\bibinfo{author}{\bibfnamefont{V.}~\bibnamefont{Ritus}},
  \bibinfo{journal}{Nucl. Phys. B} \textbf{\bibinfo{volume}{44}},
  \bibinfo{pages}{236} (\bibinfo{year}{1972}).

\bibitem[{\citenamefont{Belkacem et~al.}(1993)\citenamefont{Belkacem, Gould,
  Feinberg, Bossingham, and Meyerhof}}]{Belkacem93}
\bibinfo{author}{\bibfnamefont{A.}~\bibnamefont{Belkacem}},
  \bibinfo{author}{\bibfnamefont{H.}~\bibnamefont{Gould}},
  \bibinfo{author}{\bibfnamefont{B.}~\bibnamefont{Feinberg}},
  \bibinfo{author}{\bibfnamefont{R.}~\bibnamefont{Bossingham}},
  \bibnamefont{and} \bibinfo{author}{\bibfnamefont{W.~E.}
  \bibnamefont{Meyerhof}}, \bibinfo{journal}{Phys. Rev. Lett.}
  \textbf{\bibinfo{volume}{71}}, \bibinfo{pages}{1514} (\bibinfo{year}{1993}).

\bibitem[{\citenamefont{Belkacem et~al.}(1994)\citenamefont{Belkacem, Gould,
  Feinberg, Bossingham, and Meyerhof}}]{Belkacem94}
\bibinfo{author}{\bibfnamefont{A.}~\bibnamefont{Belkacem}},
  \bibinfo{author}{\bibfnamefont{H.}~\bibnamefont{Gould}},
  \bibinfo{author}{\bibfnamefont{B.}~\bibnamefont{Feinberg}},
  \bibinfo{author}{\bibfnamefont{R.}~\bibnamefont{Bossingham}},
  \bibnamefont{and} \bibinfo{author}{\bibfnamefont{W.~E.}
  \bibnamefont{Meyerhof}}, \bibinfo{journal}{Phys. Rev. Lett.}
  \textbf{\bibinfo{volume}{73}}, \bibinfo{pages}{2432} (\bibinfo{year}{1994}).

\bibitem[{\citenamefont{Momberger et~al.}(1987)\citenamefont{Momberger, Gr\"un,
  Scheid, Becker, and Soff}}]{Momberger87}
\bibinfo{author}{\bibfnamefont{K.}~\bibnamefont{Momberger}},
  \bibinfo{author}{\bibfnamefont{N.}~\bibnamefont{Gr\"un}},
  \bibinfo{author}{\bibfnamefont{W.}~\bibnamefont{Scheid}},
  \bibinfo{author}{\bibfnamefont{U.}~\bibnamefont{Becker}}, \bibnamefont{and}
  \bibinfo{author}{\bibfnamefont{G.}~\bibnamefont{Soff}}, \bibinfo{journal}{J.
  Phys. B} \textbf{\bibinfo{volume}{20}}, \bibinfo{pages}{L281}
  (\bibinfo{year}{1987}).

\bibitem[{\citenamefont{Yakovlev}(1966)}]{Yakovlev66}
\bibinfo{author}{\bibfnamefont{V.}~\bibnamefont{Yakovlev}},
  \bibinfo{journal}{Sov. Phys. JETP} \textbf{\bibinfo{volume}{22}},
  \bibinfo{pages}{223} (\bibinfo{year}{1966}).

\bibitem[{\citenamefont{Mittleman}(1987)}]{Mittleman87}
\bibinfo{author}{\bibfnamefont{M.~H.} \bibnamefont{Mittleman}},
  \bibinfo{journal}{Phys. Rev. A} \textbf{\bibinfo{volume}{35}},
  \bibinfo{pages}{4624} (\bibinfo{year}{1987}).

\bibitem[{\citenamefont{Roshchupkin}(1996)}]{Roshchupkin96}
\bibinfo{author}{\bibfnamefont{S.~P.} \bibnamefont{Roshchupkin}},
  \bibinfo{journal}{Laser Phys.} \textbf{\bibinfo{volume}{6}},
  \bibinfo{pages}{837} (\bibinfo{year}{1996}).

\bibitem[{\citenamefont{Dietz and Pr{\"o}bsting}(1998)}]{Dietz98}
\bibinfo{author}{\bibfnamefont{K.}~\bibnamefont{Dietz}} \bibnamefont{and}
  \bibinfo{author}{\bibfnamefont{M.}~\bibnamefont{Pr{\"o}bsting}},
  \bibinfo{journal}{J. Phys. B} \textbf{\bibinfo{volume}{31}},
  \bibinfo{pages}{L409} (\bibinfo{year}{1998}).

\bibitem[{\citenamefont{Roshchupkin}(2001)}]{Roshchupkin01}
\bibinfo{author}{\bibfnamefont{S.~P.} \bibnamefont{Roshchupkin}},
  \bibinfo{journal}{Phys. At. Nucl.} \textbf{\bibinfo{volume}{64}},
  \bibinfo{pages}{243} (\bibinfo{year}{2001}).

\bibitem[{\citenamefont{Avetissian et~al.}(2003)\citenamefont{Avetissian,
  Avetissian, Mkrtchian, and Sedrakian}}]{Avetissian03}
\bibinfo{author}{\bibfnamefont{H.~K.} \bibnamefont{Avetissian}},
  \bibinfo{author}{\bibfnamefont{A.~K.} \bibnamefont{Avetissian}},
  \bibinfo{author}{\bibfnamefont{G.~F.} \bibnamefont{Mkrtchian}},
  \bibnamefont{and} \bibinfo{author}{\bibfnamefont{K.~V.}
  \bibnamefont{Sedrakian}}, \bibinfo{journal}{Nucl. Instrum. Meth. Phys. Res.
  A} \textbf{\bibinfo{volume}{507}}, \bibinfo{pages}{582}
  (\bibinfo{year}{2003}).

\bibitem[{\citenamefont{M\"{u}ller
  et~al.}(2003{\natexlab{a}})\citenamefont{M\"{u}ller, Voitkiv, and
  Gr\"un}}]{Muller03a}
\bibinfo{author}{\bibfnamefont{C.}~\bibnamefont{M\"{u}ller}},
  \bibinfo{author}{\bibfnamefont{A.~B.} \bibnamefont{Voitkiv}},
  \bibnamefont{and} \bibinfo{author}{\bibfnamefont{N.}~\bibnamefont{Gr\"un}},
  \bibinfo{journal}{Phys. Rev. A} \textbf{\bibinfo{volume}{67}},
  \bibinfo{pages}{063407} (\bibinfo{year}{2003}{\natexlab{a}}).

\bibitem[{\citenamefont{M\"{u}ller
  et~al.}(2003{\natexlab{b}})\citenamefont{M\"{u}ller, Voitkiv, and
  Gr\"{u}n}}]{Muller03c}
\bibinfo{author}{\bibfnamefont{C.}~\bibnamefont{M\"{u}ller}},
  \bibinfo{author}{\bibfnamefont{A.~B.} \bibnamefont{Voitkiv}},
  \bibnamefont{and} \bibinfo{author}{\bibfnamefont{N.}~\bibnamefont{Gr\"{u}n}},
  \bibinfo{journal}{Nucl. Instrum. Meth. Phys. Res. B}
  \textbf{\bibinfo{volume}{205}}, \bibinfo{pages}{306}
  (\bibinfo{year}{2003}{\natexlab{b}}).

\bibitem[{\citenamefont{M\"{u}ller et~al.}(2004)\citenamefont{M\"{u}ller,
  Voitkiv, and Gr\"{u}n}}]{Muller04}
\bibinfo{author}{\bibfnamefont{C.}~\bibnamefont{M\"{u}ller}},
  \bibinfo{author}{\bibfnamefont{A.~B.} \bibnamefont{Voitkiv}},
  \bibnamefont{and} \bibinfo{author}{\bibfnamefont{N.}~\bibnamefont{Gr\"{u}n}},
  \bibinfo{journal}{Phys. Rev. A} \textbf{\bibinfo{volume}{70}},
  \bibinfo{eid}{023412} (\bibinfo{year}{2004}).

\bibitem[{\citenamefont{Kaminski et~al.}(2006)\citenamefont{Kaminski,
  Krajewska, and Ehlotzky}}]{Kaminski06}
\bibinfo{author}{\bibfnamefont{J.~Z.} \bibnamefont{Kaminski}},
  \bibinfo{author}{\bibfnamefont{K.}~\bibnamefont{Krajewska}},
  \bibnamefont{and} \bibinfo{author}{\bibfnamefont{F.}~\bibnamefont{Ehlotzky}},
  \bibinfo{journal}{Phys. Rev. A} \textbf{\bibinfo{volume}{74}},
  \bibinfo{eid}{033402} (\bibinfo{year}{2006}).

\bibitem[{\citenamefont{Krajewska et~al.}(2006)\citenamefont{Krajewska,
  Kaminski, and Ehlotzky}}]{Krajewska06}
\bibinfo{author}{\bibfnamefont{K.}~\bibnamefont{Krajewska}},
  \bibinfo{author}{\bibfnamefont{J.~Z.} \bibnamefont{Kaminski}},
  \bibnamefont{and} \bibinfo{author}{\bibfnamefont{F.}~\bibnamefont{Ehlotzky}},
  \bibinfo{journal}{Laser Phys.} \textbf{\bibinfo{volume}{16}},
  \bibinfo{pages}{272} (\bibinfo{year}{2006}).

\bibitem[{\citenamefont{Sieczka et~al.}(2006)\citenamefont{Sieczka, Krajewska,
  Kaminski, Panek, and Ehlotzky}}]{Sieczka06}
\bibinfo{author}{\bibfnamefont{P.}~\bibnamefont{Sieczka}},
  \bibinfo{author}{\bibfnamefont{K.}~\bibnamefont{Krajewska}},
  \bibinfo{author}{\bibfnamefont{J.~Z.} \bibnamefont{Kaminski}},
  \bibinfo{author}{\bibfnamefont{P.}~\bibnamefont{Panek}}, \bibnamefont{and}
  \bibinfo{author}{\bibfnamefont{F.}~\bibnamefont{Ehlotzky}},
  \bibinfo{journal}{Phys. Rev. A} \textbf{\bibinfo{volume}{73}},
  \bibinfo{pages}{053409} (\bibinfo{year}{2006}).

\bibitem[{\citenamefont{Milstein et~al.}(2006)\citenamefont{Milstein,
  M\"{u}ller, Hatsagortsyan, Jentschura, and Keitel}}]{Milstein06}
\bibinfo{author}{\bibfnamefont{A.~I.} \bibnamefont{Milstein}},
  \bibinfo{author}{\bibfnamefont{C.}~\bibnamefont{M\"{u}ller}},
  \bibinfo{author}{\bibfnamefont{K.~Z.} \bibnamefont{Hatsagortsyan}},
  \bibinfo{author}{\bibfnamefont{U.~D.} \bibnamefont{Jentschura}},
  \bibnamefont{and} \bibinfo{author}{\bibfnamefont{C.~H.}
  \bibnamefont{Keitel}}, \bibinfo{journal}{Phys. Rev. A}
  \textbf{\bibinfo{volume}{73}}, \bibinfo{eid}{062106} (\bibinfo{year}{2006}).

\bibitem[{\citenamefont{Kuchiev and Robinson}(2007)}]{Kuchiev07}
\bibinfo{author}{\bibfnamefont{M.~Y.} \bibnamefont{Kuchiev}} \bibnamefont{and}
  \bibinfo{author}{\bibfnamefont{D.~J.} \bibnamefont{Robinson}},
  \bibinfo{journal}{Phys. Rev. A} \textbf{\bibinfo{volume}{76}},
  \bibinfo{eid}{012107} (\bibinfo{year}{2007}).

\bibitem[{\citenamefont{Krajewska and Kaminski}(2008)}]{Krajewska08}
\bibinfo{author}{\bibfnamefont{K.}~\bibnamefont{Krajewska}} \bibnamefont{and}
  \bibinfo{author}{\bibfnamefont{J.~Z.} \bibnamefont{Kaminski}},
  \bibinfo{journal}{Laser Phys.} \textbf{\bibinfo{volume}{18}},
  \bibinfo{pages}{185} (\bibinfo{year}{2008}).

\bibitem[{\citenamefont{Kuchiev}(2007)}]{Kuchiev07b}
\bibinfo{author}{\bibfnamefont{M.~Y.} \bibnamefont{Kuchiev}},
  \bibinfo{journal}{Phys. Rev. Lett.} \textbf{\bibinfo{volume}{99}},
  \bibinfo{pages}{130404} (\bibinfo{year}{2007}).

\bibitem[{\citenamefont{M\"uller et~al.}(accepted)\citenamefont{M\"uller,
  Deneke, and Keitel}}]{Muller08}
\bibinfo{author}{\bibfnamefont{C.}~\bibnamefont{M\"uller}},
  \bibinfo{author}{\bibfnamefont{C.}~\bibnamefont{Deneke}}, \bibnamefont{and}
  \bibinfo{author}{\bibfnamefont{C.}~\bibnamefont{Keitel}},
  \bibinfo{journal}{Phys. Rev. Lett.}  (\bibinfo{year}{accepted}).

\bibitem[{\citenamefont{Sauter}(1931)}]{Sauter31}
\bibinfo{author}{\bibfnamefont{F.}~\bibnamefont{Sauter}},
  \bibinfo{journal}{Ann. Physik} \textbf{\bibinfo{volume}{11}},
  \bibinfo{pages}{454} (\bibinfo{year}{1931}).

\bibitem[{\citenamefont{Agger and S\o{}rensen}(1997)}]{Agger97}
\bibinfo{author}{\bibfnamefont{C.~K.} \bibnamefont{Agger}} \bibnamefont{and}
  \bibinfo{author}{\bibnamefont{S\o{}rensen}}, \bibinfo{journal}{Phys. Rev. A}
  \textbf{\bibinfo{volume}{55}}, \bibinfo{pages}{402} (\bibinfo{year}{1997}).

\bibitem[{\citenamefont{M\"{u}ller
  et~al.}(2003{\natexlab{c}})\citenamefont{M\"{u}ller, Voitkiv, and
  Gr\"un}}]{Muller03b}
\bibinfo{author}{\bibfnamefont{C.}~\bibnamefont{M\"{u}ller}},
  \bibinfo{author}{\bibfnamefont{A.~B.} \bibnamefont{Voitkiv}},
  \bibnamefont{and} \bibinfo{author}{\bibfnamefont{N.}~\bibnamefont{Gr\"un}},
  \bibinfo{journal}{Phys. Rev. Lett.} \textbf{\bibinfo{volume}{91}},
  \bibinfo{pages}{223601} (\bibinfo{year}{2003}{\natexlab{c}}).

\bibitem[{\citenamefont{Matveev et~al.}(2005)\citenamefont{Matveev, Gusarevich,
  and Pashev}}]{Matveev05}
\bibinfo{author}{\bibfnamefont{V.~I.} \bibnamefont{Matveev}},
  \bibinfo{author}{\bibfnamefont{E.~S.} \bibnamefont{Gusarevich}},
  \bibnamefont{and} \bibinfo{author}{\bibfnamefont{I.~N.}
  \bibnamefont{Pashev}}, \bibinfo{journal}{J. Exp. Theor. Phys.}
  \textbf{\bibinfo{volume}{100}}, \bibinfo{pages}{1043} (\bibinfo{year}{2005}).

\bibitem[{\citenamefont{{Di Piazza} et~al.}(2008)\citenamefont{{Di Piazza},
  Hatsagortsyan, and Keitel}}]{Piazza08}
\bibinfo{author}{\bibfnamefont{A.}~\bibnamefont{{Di Piazza}}},
  \bibinfo{author}{\bibfnamefont{K.~Z.} \bibnamefont{Hatsagortsyan}},
  \bibnamefont{and} \bibinfo{author}{\bibfnamefont{C.~H.}
  \bibnamefont{Keitel}}, \bibinfo{journal}{Phys. Rev. Lett.}
  \textbf{\bibinfo{volume}{100}}, \bibinfo{pages}{010403}
  (\bibinfo{year}{2008}).

\bibitem[{\citenamefont{Sch\"utzhold et~al.}(2006)\citenamefont{Sch\"utzhold,
  Schaller, and Habs}}]{Schutzhold06}
\bibinfo{author}{\bibfnamefont{R.}~\bibnamefont{Sch\"utzhold}},
  \bibinfo{author}{\bibfnamefont{G.}~\bibnamefont{Schaller}}, \bibnamefont{and}
  \bibinfo{author}{\bibfnamefont{D.}~\bibnamefont{Habs}},
  \bibinfo{journal}{Phys. Rev. Lett.} \textbf{\bibinfo{volume}{97}},
  \bibinfo{pages}{121302} (\bibinfo{year}{2006}).

\bibitem[{\citenamefont{Sch\"utzhold et~al.}(2008)\citenamefont{Sch\"utzhold,
  Schaller, and Habs}}]{Schutzhold08}
\bibinfo{author}{\bibfnamefont{R.}~\bibnamefont{Sch\"utzhold}},
  \bibinfo{author}{\bibfnamefont{G.}~\bibnamefont{Schaller}}, \bibnamefont{and}
  \bibinfo{author}{\bibfnamefont{D.}~\bibnamefont{Habs}},
  \bibinfo{journal}{Phys. Rev. Lett.} \textbf{\bibinfo{volume}{100}},
  \bibinfo{pages}{091301} (\bibinfo{year}{2008}).

\bibitem[{\citenamefont{{Di Piazza} et~al.}(2005)\citenamefont{{Di Piazza},
  Hatsagortsyan, and Keitel}}]{Piazza05}
\bibinfo{author}{\bibfnamefont{A.}~\bibnamefont{{Di Piazza}}},
  \bibinfo{author}{\bibfnamefont{K.~Z.} \bibnamefont{Hatsagortsyan}},
  \bibnamefont{and} \bibinfo{author}{\bibfnamefont{C.~H.}
  \bibnamefont{Keitel}}, \bibinfo{journal}{Phys. Rev. D}
  \textbf{\bibinfo{volume}{72}}, \bibinfo{pages}{085005}
  (\bibinfo{year}{2005}).

\bibitem[{\citenamefont{Lundstr\"om et~al.}(2006)\citenamefont{Lundstr\"om,
  Brodin, Lundin, Marklund, Bingham, Collier, Mendon\c{c}a, and
  Norreys}}]{Lundstrom06}
\bibinfo{author}{\bibfnamefont{E.}~\bibnamefont{Lundstr\"om}},
  \bibinfo{author}{\bibfnamefont{G.}~\bibnamefont{Brodin}},
  \bibinfo{author}{\bibfnamefont{J.}~\bibnamefont{Lundin}},
  \bibinfo{author}{\bibfnamefont{M.}~\bibnamefont{Marklund}},
  \bibinfo{author}{\bibfnamefont{R.}~\bibnamefont{Bingham}},
  \bibinfo{author}{\bibfnamefont{J.}~\bibnamefont{Collier}},
  \bibinfo{author}{\bibfnamefont{J.~T.} \bibnamefont{Mendon\c{c}a}},
  \bibnamefont{and} \bibinfo{author}{\bibfnamefont{P.}~\bibnamefont{Norreys}},
  \bibinfo{journal}{Phys. Rev. Lett.} \textbf{\bibinfo{volume}{96}},
  \bibinfo{pages}{083602} (\bibinfo{year}{2006}).

\bibitem[{\citenamefont{Fedotov and Narozhny}(2007)}]{Fedotov07}
\bibinfo{author}{\bibfnamefont{A.}~\bibnamefont{Fedotov}} \bibnamefont{and}
  \bibinfo{author}{\bibfnamefont{N.}~\bibnamefont{Narozhny}},
  \bibinfo{journal}{Phys. Lett. A} \textbf{\bibinfo{volume}{362}},
  \bibinfo{pages}{1} (\bibinfo{year}{2007}).

\bibitem[{\citenamefont{{Di Piazza} et~al.}(2007)\citenamefont{{Di Piazza},
  Milstein, and Keitel}}]{Piazza07}
\bibinfo{author}{\bibfnamefont{A.}~\bibnamefont{{Di Piazza}}},
  \bibinfo{author}{\bibfnamefont{A.~I.} \bibnamefont{Milstein}},
  \bibnamefont{and} \bibinfo{author}{\bibfnamefont{C.~H.}
  \bibnamefont{Keitel}}, \bibinfo{journal}{Phys. Rev. A}
  \textbf{\bibinfo{volume}{76}}, \bibinfo{pages}{032103}
  (\bibinfo{year}{2007}).

\bibitem[{\citenamefont{Brodin et~al.}(2007)\citenamefont{Brodin, Marklund,
  Eliasson, and Shukla}}]{Brodin07}
\bibinfo{author}{\bibfnamefont{G.}~\bibnamefont{Brodin}},
  \bibinfo{author}{\bibfnamefont{M.}~\bibnamefont{Marklund}},
  \bibinfo{author}{\bibfnamefont{B.}~\bibnamefont{Eliasson}}, \bibnamefont{and}
  \bibinfo{author}{\bibfnamefont{P.~K.} \bibnamefont{Shukla}},
  \bibinfo{journal}{Phys. Rev. Lett.} \textbf{\bibinfo{volume}{98}},
  \bibinfo{pages}{125001} (\bibinfo{year}{2007}).

\bibitem[{\citenamefont{{Di Piazza} et~al.}(2006)\citenamefont{{Di Piazza},
  Hatsagortsyan, and Keitel}}]{Piazza06}
\bibinfo{author}{\bibfnamefont{A.}~\bibnamefont{{Di Piazza}}},
  \bibinfo{author}{\bibfnamefont{K.~Z.} \bibnamefont{Hatsagortsyan}},
  \bibnamefont{and} \bibinfo{author}{\bibfnamefont{C.~H.}
  \bibnamefont{Keitel}}, \bibinfo{journal}{Phys. Rev. Lett.}
  \textbf{\bibinfo{volume}{97}}, \bibinfo{pages}{083603}
  (\bibinfo{year}{2006}).

\bibitem[{\citenamefont{Heinzl et~al.}(2006)\citenamefont{Heinzl, Liesfeld,
  Amthor, Schwoerer, Sauerbrey, and Wipf}}]{Heinzl06}
\bibinfo{author}{\bibfnamefont{T.}~\bibnamefont{Heinzl}},
  \bibinfo{author}{\bibfnamefont{B.}~\bibnamefont{Liesfeld}},
  \bibinfo{author}{\bibfnamefont{K.-U.} \bibnamefont{Amthor}},
  \bibinfo{author}{\bibfnamefont{H.}~\bibnamefont{Schwoerer}},
  \bibinfo{author}{\bibfnamefont{R.}~\bibnamefont{Sauerbrey}},
  \bibnamefont{and} \bibinfo{author}{\bibfnamefont{A.}~\bibnamefont{Wipf}},
  \bibinfo{journal}{Opt. Commun.} \textbf{\bibinfo{volume}{267}},
  \bibinfo{pages}{318} (\bibinfo{year}{2006}).

\bibitem[{\citenamefont{Krekora et~al.}(2005)\citenamefont{Krekora, Cooley, Su,
  and Grobe}}]{Krekora05}
\bibinfo{author}{\bibfnamefont{P.}~\bibnamefont{Krekora}},
  \bibinfo{author}{\bibfnamefont{K.}~\bibnamefont{Cooley}},
  \bibinfo{author}{\bibfnamefont{Q.}~\bibnamefont{Su}}, \bibnamefont{and}
  \bibinfo{author}{\bibfnamefont{R.}~\bibnamefont{Grobe}},
  \bibinfo{journal}{Phys. Rev. Lett.} \textbf{\bibinfo{volume}{95}},
  \bibinfo{pages}{070403} (\bibinfo{year}{2005}).

\bibitem[{\citenamefont{Bjorken and Drell}(1964)}]{Bjorken64}
\bibinfo{author}{\bibfnamefont{J.~D.} \bibnamefont{Bjorken}} \bibnamefont{and}
  \bibinfo{author}{\bibfnamefont{S.~D.} \bibnamefont{Drell}},
  \emph{\bibinfo{title}{Relativistische Quantenmechanik}}
  (\bibinfo{publisher}{Bibliogr. Inst., Mannheim}, \bibinfo{year}{1964}).

\bibitem[{\citenamefont{Joachain et~al.}(2000)\citenamefont{Joachain, D\"orr,
  and Kylstra}}]{Joachain00}
\bibinfo{author}{\bibfnamefont{C.}~\bibnamefont{Joachain}},
  \bibinfo{author}{\bibfnamefont{M.}~\bibnamefont{D\"orr}}, \bibnamefont{and}
  \bibinfo{author}{\bibfnamefont{N.}~\bibnamefont{Kylstra}},
  \bibinfo{journal}{Adv. At. Mol. Phys.} \textbf{\bibinfo{volume}{42}},
  \bibinfo{pages}{225} (\bibinfo{year}{2000}).

\bibitem[{\citenamefont{{Becker et al.}}(2002)}]{Becker02}
\bibinfo{author}{\bibfnamefont{W.}~\bibnamefont{{Becker et al.}}},
  \bibinfo{journal}{Adv. At. Mol. Opt. Phys.} \textbf{\bibinfo{volume}{48}},
  \bibinfo{pages}{35} (\bibinfo{year}{2002}).

\bibitem[{\citenamefont{Milo\v{s}evi\'c and Ehlotzky}(2003)}]{Milosevic03}
\bibinfo{author}{\bibfnamefont{D.~B.} \bibnamefont{Milo\v{s}evi\'c}}
  \bibnamefont{and} \bibinfo{author}{\bibfnamefont{F.}~\bibnamefont{Ehlotzky}},
  \bibinfo{journal}{Adv. At. Mol. Opt. Phys.} \textbf{\bibinfo{volume}{49}},
  \bibinfo{pages}{373} (\bibinfo{year}{2003}).

\bibitem[{\citenamefont{Reiss}(1990{\natexlab{a}})}]{Reiss90}
\bibinfo{author}{\bibfnamefont{H.~R.} \bibnamefont{Reiss}},
  \bibinfo{journal}{J. Opt. Soc. Am. B} \textbf{\bibinfo{volume}{7}},
  \bibinfo{pages}{574} (\bibinfo{year}{1990}{\natexlab{a}}).

\bibitem[{\citenamefont{Crawford and Reiss}(1998)}]{Crawford98}
\bibinfo{author}{\bibfnamefont{D.}~\bibnamefont{Crawford}} \bibnamefont{and}
  \bibinfo{author}{\bibfnamefont{H.~R.} \bibnamefont{Reiss}},
  \bibinfo{journal}{Opt. Express} \textbf{\bibinfo{volume}{2}},
  \bibinfo{pages}{289} (\bibinfo{year}{1998}).

\bibitem[{\citenamefont{Keldysh}(1965)}]{Keldysh65}
\bibinfo{author}{\bibfnamefont{L.~V.} \bibnamefont{Keldysh}},
  \bibinfo{journal}{Sov. Phys. JETP} \textbf{\bibinfo{volume}{20}},
  \bibinfo{pages}{1307} (\bibinfo{year}{1965}).

\bibitem[{\citenamefont{Faisal}(1973)}]{Faisal73}
\bibinfo{author}{\bibfnamefont{F.~H.~M.} \bibnamefont{Faisal}},
  \bibinfo{journal}{J. Phys. B} \textbf{\bibinfo{volume}{6}},
  \bibinfo{pages}{L89} (\bibinfo{year}{1973}).

\bibitem[{\citenamefont{Reiss}(1980)}]{Reiss80b}
\bibinfo{author}{\bibfnamefont{H.~R.} \bibnamefont{Reiss}},
  \bibinfo{journal}{Phys. Rev. A} \textbf{\bibinfo{volume}{22}},
  \bibinfo{pages}{1786} (\bibinfo{year}{1980}).

\bibitem[{\citenamefont{Reiss}(1990{\natexlab{b}})}]{Reiss90b}
\bibinfo{author}{\bibfnamefont{H.~R.} \bibnamefont{Reiss}},
  \bibinfo{journal}{Phys. Rev. A} \textbf{\bibinfo{volume}{42}},
  \bibinfo{pages}{1476} (\bibinfo{year}{1990}{\natexlab{b}}).

\bibitem[{\citenamefont{Reiss}(1992)}]{Reiss92}
\bibinfo{author}{\bibfnamefont{H.~R.} \bibnamefont{Reiss}},
  \bibinfo{journal}{Prog. Quantum Electron.} \textbf{\bibinfo{volume}{16}},
  \bibinfo{pages}{1} (\bibinfo{year}{1992}).

\bibitem[{\citenamefont{Volkov}(1935)}]{Volkov35}
\bibinfo{author}{\bibfnamefont{D.~M.} \bibnamefont{Volkov}},
  \bibinfo{journal}{Z. Phys.} \textbf{\bibinfo{volume}{94}},
  \bibinfo{pages}{250} (\bibinfo{year}{1935}).

\bibitem[{\citenamefont{Beresteckij et~al.}(1991)\citenamefont{Beresteckij,
  Lifsic, and Pitaevskij}}]{Landau91}
\bibinfo{author}{\bibfnamefont{V.~B.} \bibnamefont{Beresteckij}},
  \bibinfo{author}{\bibfnamefont{E.~M.} \bibnamefont{Lifsic}},
  \bibnamefont{and} \bibinfo{author}{\bibfnamefont{L.~P.}
  \bibnamefont{Pitaevskij}}, \emph{\bibinfo{title}{Quantenelektrodynamik}},
  Lehrbuch der theoretischen Physik / L. D. Landau ; E. M. Lifschitz ; 4
  (\bibinfo{publisher}{Akad.-Verl., Frankfurt am Main}, \bibinfo{year}{1991}).

\bibitem[{\citenamefont{Kopold et~al.}(2002)\citenamefont{Kopold, Becker,
  Kleber, and Paulus}}]{Kopold02}
\bibinfo{author}{\bibfnamefont{R.}~\bibnamefont{Kopold}},
  \bibinfo{author}{\bibfnamefont{W.}~\bibnamefont{Becker}},
  \bibinfo{author}{\bibfnamefont{M.}~\bibnamefont{Kleber}}, \bibnamefont{and}
  \bibinfo{author}{\bibfnamefont{G.~G.} \bibnamefont{Paulus}},
  \bibinfo{journal}{J. Phys. B} \textbf{\bibinfo{volume}{35}},
  \bibinfo{pages}{217} (\bibinfo{year}{2002}).

\bibitem[{\citenamefont{Pratt}(1960)}]{Pratt60b}
\bibinfo{author}{\bibfnamefont{R.~H.} \bibnamefont{Pratt}},
  \bibinfo{journal}{Phys. Rev.} \textbf{\bibinfo{volume}{119}},
  \bibinfo{pages}{1619} (\bibinfo{year}{1960}).

\bibitem[{\citenamefont{Uggerhoj}(2006)}]{Uggerhoj06}
\bibinfo{author}{\bibfnamefont{U.~I.} \bibnamefont{Uggerhoj}},
  \bibinfo{journal}{Phys. Rev. A} \textbf{\bibinfo{volume}{73}},
  \bibinfo{eid}{052705} (\bibinfo{year}{2006}).

\end{thebibliography}


\end{document}